\documentclass[11pt]{article}

\usepackage[preprint]{acl}

\usepackage{times}
\usepackage{latexsym}

\usepackage[T1]{fontenc}

\usepackage[utf8]{inputenc}

\usepackage{microtype}

\usepackage{inconsolata}

\usepackage{graphicx}
\usepackage{booktabs}
\usepackage{multirow}
\usepackage{colortbl}

\usepackage{algorithm}
\usepackage{algpseudocode}
\usepackage{amsmath,amssymb}
\usepackage{amsfonts}
\algrenewcommand{\algorithmicindent}{1.0em} 

\usepackage{enumitem}
\usepackage[table]{xcolor}

\usepackage{tcolorbox}
\tcbuselibrary{breakable,listings}
\usepackage{fvextra}
\usepackage{xcolor}

\DeclareCaptionType{prompt}[Prompt][List of Prompts]
\title{SynWeaver: Website-Prior Task and Trajectory Co-Synthesis for Web Agents}

\author{
  \textbf{Ruitao Wang} \and
  \textbf{Yuwen Hao} \and
  \textbf{Menglin Yang}\textsuperscript{\textdagger} \\
  Hong Kong University of Science and Technology (Guangzhou) \\
  \texttt{\{rwang356, yhao481\}@connect.hkust-gz.edu.cn} \\
  \texttt{menglinyang@hkust-gz.edu.cn} \\
  {\small\textsuperscript{\textdagger}Corresponding author.}
}

\begin{document}
\maketitle

\begin{abstract}
Web agents often struggle to generalize to unseen websites because they lack website-specific supervision. Recent exploration-based data synthesis methods reduce manual annotation, but they still face two key limitations: they often fail to cover the full functionality of a website, and without sufficient website prior knowledge, they tend to propose hallucinated tasks, which in turn limits the diversity and efficiency of downstream trajectory synthesis. We present \textbf{SynWeaver}, a website-prior task-trajectory co-synthesis framework designed to address these challenges. SynWeaver first performs structured website exploration and constructs a website map that covers a broad set of functionally distinct page states and executable interactions on the target website. It then derives page-level and transition-level supervision from this map to train a UI-aware model with website-specific priors, enabling more grounded task proposals. Finally, SynWeaver performs collaborative task-trajectory synthesis, jointly updating the task and execution trajectory when they become inconsistent, and then verifies and repairs the collected results to produce executable, semantically aligned supervision. Experiments on WebArena and WebVoyager demonstrate that SynWeaver consistently outperforms strong synthesis baselines and yields more effective supervision for both in-domain and out-of-domain generalization. The code is publicly available at \url{https://github.com/Eilok/SynWeaver}.
\end{abstract}

\section{Introduction}

End-to-end web agents have shown strong potential on realistic tasks, but training effective agents still requires substantial task-trajectory supervision \citep{qin2025ui, wang2025ui}. Obtaining such data at scale is challenging because manual trajectory annotation is costly and time-consuming \citep{deng2023mind2web}. Moreover, even with interaction data, agents often transfer poorly to unseen websites, as multimodal models lack website-specific priors about page structure, interaction affordances, and action consequences \citep{tang2025survey}. Therefore, web-agent data synthesis is important not only for alleviating the supervision bottleneck, but also for injecting target-website knowledge that improves generalization.

Existing web agent data synthesis methods can be broadly divided into two families. The first family leverages external resources, such as tutorials or documentation, to instantiate tasks and replay trajectories \citep{zhang2026tongui,xu2025agenttrek,su2025learn,shen2024scribeagent}. These methods can provide useful supervision when such resources are abundant, but their coverage is fundamentally bounded by what already exists in the internet. For many websites, relevant materials are sparse, outdated, or entirely unavailable \citep{mereu2026specializing}. In this work, we instead focus on the second family, which mines tasks directly from website interaction and is therefore applicable even when external artifacts are missing \citep{sun2025genesis,pahuja2025explorer, gandhi2025go}.

Within this family, a central challenge is generating tasks that are both realistic and executable. Some methods synthesize tasks from local interactions, which is efficient but often short-sighted: lacking sufficient website-specific knowledge \citep{shi2025gui}, the model may propose tasks that appear plausible on the current page yet become infeasible during subsequent interaction. NNetNav \citep{murty2024nnetnav} alleviates this issue through multi-step LLM exploration, but at the cost of substantial redundant interaction. SynthAgent \citep{wang2025adapting} further introduces task refinement during execution and trajectory refinement after execution, partially mitigating infeasible initial tasks. However, because these refinements are decoupled, the task is often rewritten to fit the collected trajectory, producing step‑heavy descriptions that read like procedural instructions rather than natural user requests. To address these limitations, we propose \textbf{SynWeaver}, a website-prior task and trajectory co-synthesis framework.

SynWeaver places explicit website knowledge at the center of synthesis and proceeds in three stages. \textit{First}, it constructs a website map of functionally distinct states and executable transitions, improving coverage while reducing redundant exploration. \textit{Second}, it derives user interface (UI) supervision from the map to fine-tune a UI-aware model, enabling task proposal grounded in website-specific priors rather than local observations alone. \textit{Third}, it performs collaborative task-trajectory synthesis, where teacher model jointly refines the task and execution prefix whenever intent and execution become inconsistent. This collaborative refinement preserves the natural form of user requests while maintaining execution feasibility.

Experiments on WebArena and WebVoyager confirm the effectiveness of SynWeaver. Using only 822 validated task-trajectory pairs synthesized from website maps, SynWeaver achieves the best overall WebArena success rate on both Qwen3-VL-8B-Instruct and InternVL3-8B, and reaches 27.06 success rate on WebVoyager, outperforming the strongest baseline by 4.64 points.

Our contributions are threefold:

\begin{enumerate}[noitemsep, topsep=0pt]
    \item We introduce a structured website exploration mechanism that reduces redundant interactions while preserving coverage of functionally distinct states, compared to random-walk exploration.
    \item We propose SynWeaver, a website-prior-centered framework for collaborative task-trajectory synthesis that couples grounded task proposal with joint refinement of task intent and execution traces.
    \item Extensive experiments and ablations show that SynWeaver produces more effective and data-efficient supervision for web agents, yielding strong in-domain and out-of-domain gains over prior synthesis methods.
\end{enumerate}
\section{Related Work}
\label{sec:related-work}

\paragraph{Executable Task Synthesis.}
Existing synthesis pipelines usually acquire environment knowledge before generating grounded tasks. Some methods mine executable tasks from online tutorials \citep{zhang2026tongui, xu2025agenttrek, shen2024scribeagent, zhou2025proposer}; Learn-by-interact further augments such seeds with Self-Instruct \citep{wang2023self, su2025learn}. Recent web-centric methods instead interact with target websites: OS-Genesis samples random-walk actions \citep{sun2025genesis}, SynthAgent reduces redundancy via function categorization \citep{wang2025adapting}, and Go-Browse builds an exploration graph to discover pages and induce tasks \citep{gandhi2025go}. However, these approaches still provide limited guarantees of functional coverage. Graph-based systematic exploration has addressed similar coverage issues in apps, where KG-RAG and GUI-Xplore use DroidBot for exhaustive exploration \citep{guan2025kg, sun2025gui, li2017droidbot}; earlier web research also uses Crawljax to traverse website states with depth-first search (DFS) \citep{mesbah2012crawling, peng2012graph}. Inspired by graph-structured web exploration \citep{chen2025pg, zhang2026webnavigator}, we construct a website map covering functionally distinct interactions and use it as an LLM fine-tuning prior, enabling task synthesis with richer website context.

\paragraph{Trajectory Data Synthesis for Web Agents.}
Web agent training requires trajectory-level supervision \citep{nguyen2025gui}, and synthetic trajectories often rely on stronger LLMs for labeling \citep{tang2025survey}. Existing methods filter tutorials with FastText before LLM annotation \citep{xu2025retrieval, bojanowski2017enriching}, roll out trajectories with learned web-world models \citep{gao2025websynthesis}, or create verifiable synthetic websites to collect interaction traces \citep{wu2026autowebworld}. For LLM demonstrations on real websites, early approaches retain only trajectories passing binary filters \citep{lin2026ui, lu2026structured}, while OS-Genesis scores trajectories for weighted training instead of discarding low-quality data \citep{sun2025genesis}. Explorer refines tasks during execution \citep{pahuja2025explorer}, and SynthAgent adds post-execution trajectory refinement \citep{wang2025adapting}. Yet decoupled task and trajectory refinement cannot reliably repair execution-time failures, reducing data efficiency and task-trajectory alignment. We therefore introduce collaborative refinement, which jointly optimizes tasks and trajectories throughout synthesis to preserve their semantic consistency.

\section{Method}

\paragraph{Problem Formulation.}
We model a website as a partially observable web environment $\mathcal{W}$. At interaction step $t$, the agent receives an observation $o_t \in \mathcal{O}$ (e.g., textual and visual content), makes a reasoning $r_t \in \mathcal{R}$ and performs an action $a_t \in \mathcal{A}$, yielding an execution history $h=(o_1,r_1,a_1,\dots,o_H,r_H,a_H)$ of length $H$. We formulate web agent data synthesis as two coupled generation processes. \textbf{Task synthesis} generates a task instruction $x$ from a partial context $h^c$, i.e., $x \sim p_\theta(x \mid h^c)$, where $h^c$ may be a local page state or an exploration prefix ending at the current state. \textbf{Trajectory synthesis} then generates an execution trajectory $h^e$ conditioned on $x$ by sequentially selecting actions according to $a_t \sim \pi(a \mid o_t,x,h_{<t})$ until termination. Our goal is to synthesize a dataset $\mathcal{D}=\{(x_i,h_i^e)\}_{i=1}^N$ of $N$ executable and semantically aligned task-trajectory pairs, rather than trajectories in isolation. In other words, the task and the trajectory are treated as co-equal objects that must express the same underlying intent.

\begin{figure*}[t]
    \centering
    \includegraphics[width=\textwidth]{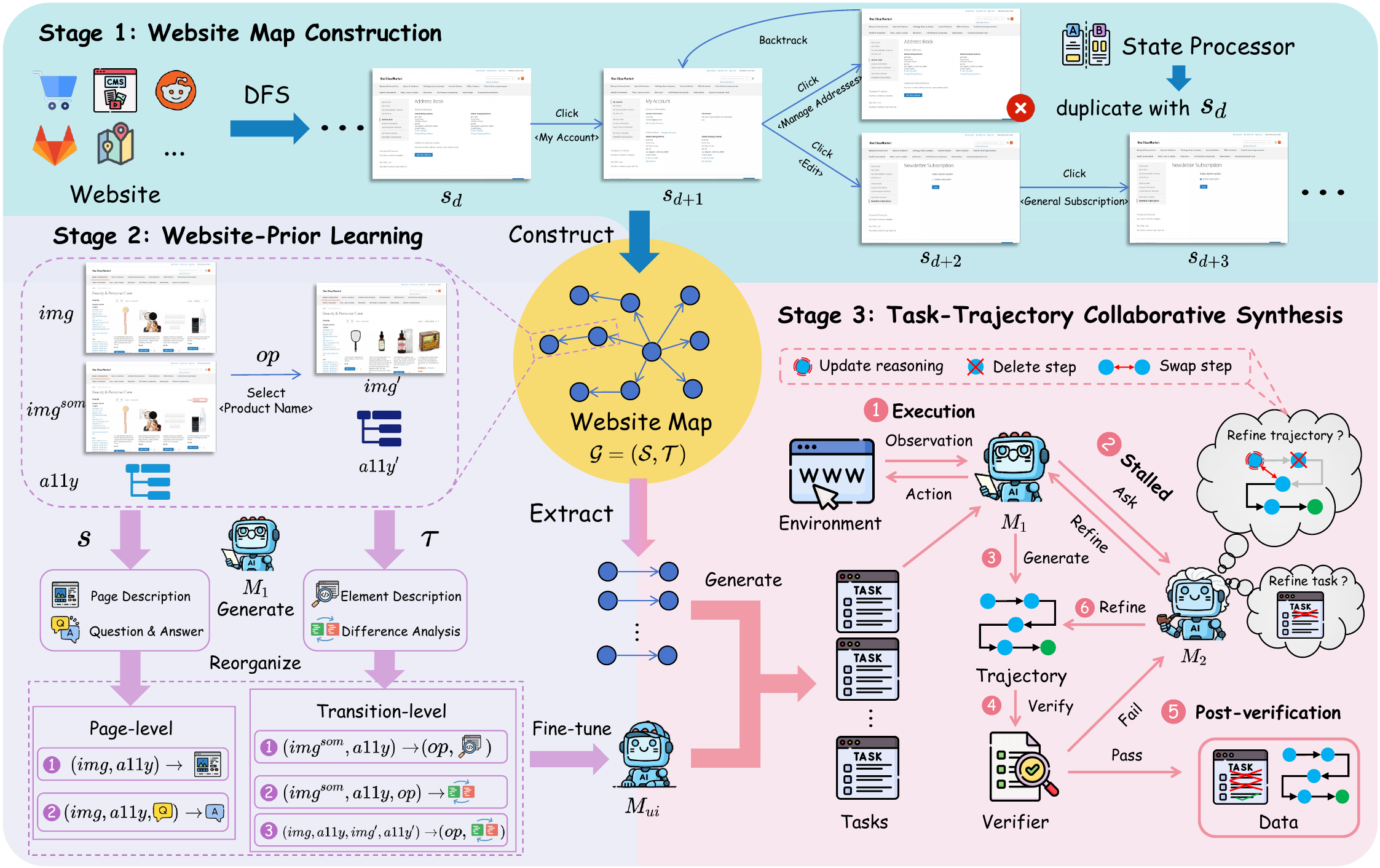}
    \caption{Overview of SynWeaver. The framework first constructs a website map, then learns website-specific UI priors from the collected states and transitions, and finally performs task-trajectory collaborative synthesis to produce executable and semantically aligned task-trajectory pairs.}
    \label{fig:method_overview}
\end{figure*}

As shown in Figure~\ref{fig:method_overview}, our SynWeaver framework proceeds in three stages: website map construction, website-prior learning, and task-trajectory collaborative synthesis. We describe each stage in detail in the following.

\subsection{Website Map Construction}
Existing works often explore websites via random walks \citep{sun2025genesis} or LLM-guided probing \citep{murty2024nnetnav}, but such strategies may repeatedly traverse elements with the same functionality while missing interactions with distinct effects, yielding redundant and incomplete traces that limit downstream task diversity. We therefore introduce a DFS-based crawler that constructs a website map $\mathcal{G}=(\mathcal{S}, \mathcal{T})$ for each target website, where each node $s \in \mathcal{S}$ denotes a functionally distinct state together with its page screenshot and accessibility tree, and each edge $\tau \in \mathcal{T}$ denotes an executable interaction and records the corresponding action. Rather than enumerating superficial page variations, $\mathcal{G}$ captures the website's functional topology, reducing redundant exploration and providing the foundation for subsequent website-prior learning and task-trajectory synthesis.

We construct $\mathcal{G}$ through depth-first exploration from a seed URL. At each visited state, the crawler enumerates executable candidates, performs the corresponding interactions, and records the resulting transitions. To avoid over-exploring superficial page variations, we introduce a unified \textit{state processor} with two functions: progressive state comparison, which determines whether a post-action page should be treated as a new state, and duplicate-trigger detection, which suppresses interactions that have already been triggered. DFS recurses only when a new state is confirmed; once a branch is exhausted, the crawler restores the previous context and resumes the remaining unexplored interactions. Full construction details are provided in Appendix~\ref{sec:crawler}.

\subsection{Website-Prior Learning} 
Existing works often propose tasks from local states \citep{sun2025genesis, wang2025adapting, pahuja2025explorer} or a short interaction prefix \citep{murty2024nnetnav}. Without sufficient website knowledge, however, the model may overfit local cues and propose tasks that are overly narrow or even infeasible. We therefore learn website-specific UI priors from the website map before task synthesis.

\paragraph{Supervision Signals.} We design five supervision formats, grouped into page-level and transition-level supervision. Let $q$, $img$, $a11y$, $img^{som}$, $op$, and $(img', a11y')$ denote the question, current screenshot, accessibility tree, SoM-annotated screenshot \citep{yang2023set}, executed action, and post-interaction state, respectively. Page-level supervision contains two forms: \textbf{page description}, which takes $(img, a11y)$ as input and predicts a detailed description of the page content, layout, and style to provide a holistic understanding of the page; and \textbf{page question answering} (QA), which takes $(img, a11y, q)$ as input and predicts the answer to an information-seeking question, thereby encouraging targeted information extraction. Transition-level supervision contains three forms: \textbf{element description}, which takes $(img^{som}, a11y)$ as input and predicts how the marked element should be interacted with and what function it serves, thereby grounding element-level affordances; \textbf{forward transition description}, which takes $(img^{som}, a11y, op)$ as input and predicts how the page changes after executing $op$, thereby modeling action-conditioned state changes; and \textbf{inverse transition description}, which takes $(img, a11y, img', a11y')$ as input and predicts a description of the observed change together with the action most likely to have caused it, thereby modeling change summarization and action inference. The data examples are shown in Appendix~\ref{sec:ui-data}.

\paragraph{Data Construction.} We construct the supervision data by querying a teacher model $M_1$ over page states and recorded transitions in the website map. For each node, one call to $M_1$ with $(img, a11y)$ yields a page description and one QA pair, which are then reorganized into page-description data and page-QA instances. For each recorded transition, one teacher call with $(img^{som}, a11y, op, img', a11y')$ yields an element function description and a transition change analysis, which are then recombined into element description, forward transition, and inverse transition supervision.

\paragraph{Training.}
We aggregate all supervision sources into a unified dataset $\mathcal{D}_{ui}$ and shuffle it. We then fine-tune the target model $M$ with LoRA \citep{hu2022lora} on $\mathcal{D}_{ui}$ to obtain a UI-aware model $M_{ui}$ that captures website-specific UI knowledge.

\subsection{Task-Trajectory Collaborative Synthesis}

\begin{figure*}[t]
    \centering
    \includegraphics[width=\textwidth]{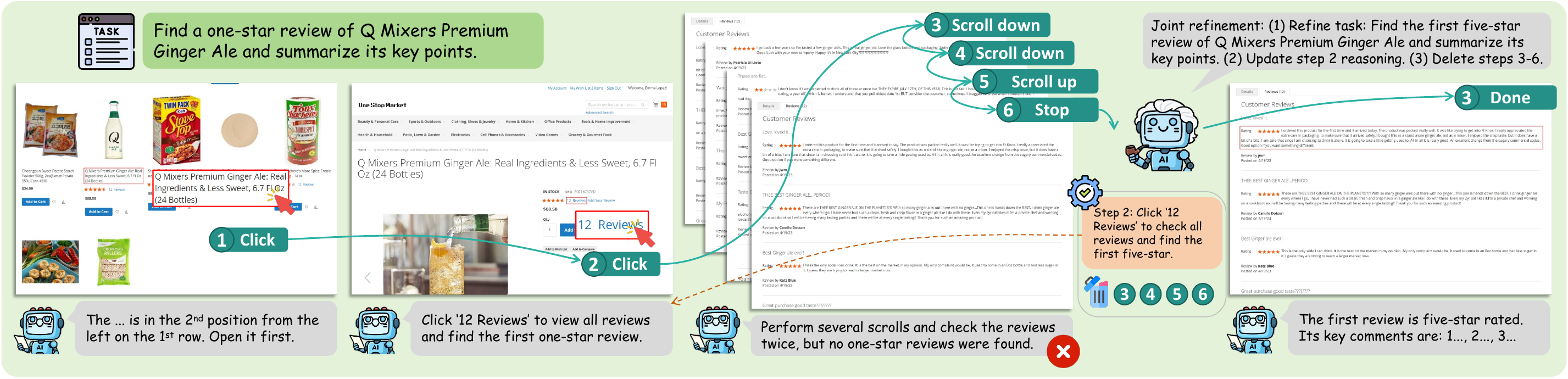}
    \caption{Case study of collaborative trajectory synthesis. When the target review is non-existent, SynWeaver adopts joint mode to refine the task and trajectory, ensuring the task can be successfully completed.}
    \label{fig:CR-case}
\end{figure*}

Prior work such as SynthAgent \citep{wang2025adapting} largely treats task refinement and trajectory refinement as separate optimization problems. This decoupling can be problematic: if one only optimizes the task, the task description may gradually align to the collected trace, yielding procedure‑heavy phrasing instead of a natural user request; if one only optimizes the trajectory against a fixed task, partially useful traces may be discarded once they no longer fit the specification, reducing data efficiency. We therefore synthesize tasks and trajectories collaboratively so that intent alignment and executability are maintained throughout data collection.

\paragraph{Reverse Task Synthesis.} We follow the reverse task synthesis strategy of OS-Genesis \citep{sun2025genesis} by extracting transition triplets $(s, op, s')$ from the website map and prompting $M_{ui}$ to infer a high-level task $x_0$ for which the observed transition is a plausible intermediate consequence. Unlike OS-Genesis, however, $M_{ui}$ already encodes website-specific UI priors, allowing it to propose tasks that are better grounded in verified website behavior and less prone to hallucinated affordances. The synthesized task $x_0$ then serves as the initial task specification.

\paragraph{Collaborative Trajectory Synthesis.} Let $x_t$ denote the task specification at step $t$, and let $h_t=(o_1, r_1, a_1, \dots, o_t, r_t, a_t)$ denote the executed prefix up to that point. Starting from $x_0$, a teacher model $M_1$ interacts with the website online to collect a trajectory. We trigger collaborative refinement whenever the current task lacks essential details, becomes incompatible with the observed website state, or execution stalls after multiple unsuccessful attempts. In such cases, we invoke a stronger teacher $M_2$ to co-optimize the task and the collected prefix. We consider two refinement modes. In the \textit{task-only} mode, $M_2$ applies $\operatorname{Update}(x_t)$ while keeping $h_t$ fixed, producing a revised task that better matches the current observations and can be completed within the next 2-3 steps. In the \textit{joint} mode, used when task-only refinement would preserve an excessively noisy prefix, $M_2$ edits both $x_t$ and $h_t$. Following SynthAgent \citep{wang2025adapting}, we allow two trajectory edits over step indices $i$ and $j$: $\operatorname{Delete}(i)$ removes an obsolete or redundant step, and $\operatorname{Reorder}(i,j)$ swaps two locally commutable steps. In addition, we introduce $\operatorname{Update}(r_t)$ to revise affected reasoning so that the refined execution remains logically coherent. Execution then resumes under the refined pair until termination, yielding a candidate trajectory $h^e$. A case is shown in Figure~\ref{fig:CR-case}.

\paragraph{Post-Verification.} Although co-optimization improves alignment during collection, some candidate trajectories may still contain failed attempts, redundant operations, or improper termination. We therefore perform post-verification after execution with access to the final task and the full trajectory. The first stage is heuristic verification, which applies three programmatic checks: (1) \textit{termination effectiveness}, requiring an explicit completion action; (2) \textit{trajectory validity}, checking successful completion within the predefined step budget; and (3) \textit{trajectory consistency}, detecting invalid repetitive action patterns. Additional implementation details are deferred to Appendix~\ref{sec:pos-val}. Pairs that pass are accepted directly. Otherwise, they enter a reconstruction stage, where $M_2$ performs targeted repair under the same \textit{joint} mode as above by applying $\operatorname{Update}(x_t)$, $\operatorname{Delete}(i)$, $\operatorname{Reorder}(i,j)$, and $\operatorname{Update}(r_t)$ as needed; we additionally allow $\operatorname{Add}(\texttt{none})$ when an explicit completion action is missing. The repaired trajectory is then returned to heuristic verification, and the reconstruction-verification loop repeats until the pair passes. If $M_2$ cannot reconstruct the data, the pair is filtered out. Validated pairs are retained in $\mathcal{D}$ as semantically aligned task-trajectory supervision.

\paragraph{Dataset Expansion.} After generating an initial set of tasks from all extracted transitions, we resample transitions and prepend previously synthesized tasks associated with the same transition to the prompt, then ask $M_{ui}$ to propose an additional task that is distinct from the existing ones. Repeating the above synthesis and verification process further expands the final dataset $\mathcal{D}$. Further details are shown in Appendix~\ref{sec:data-expand}. 
\section{Experiments}

\definecolor{ourslavender}{RGB}{245,239,255}

\subsection{Experimental Setup}

We evaluate SynWeaver on standard web agent benchmarks under matched fine-tuning settings.

\paragraph{Benchmarks.} We evaluate on WebArena \citep{zhou2024webarena}, a self-hostable benchmark spanning five websites, and WebVoyager \citep{he2024webvoyager}, an online benchmark covering diverse real-world websites. Following prior work \citep{wang2025adapting}, we sample one task per WebArena template and remove cross-website tasks, yielding 226 evaluation tasks. For WebVoyager, we apply executability and overlap filtering, leaving 388 evaluation tasks. We report task success rate (SR) under the official protocols. SynWeaver synthesizes data on the five WebArena websites and is evaluated in-domain on WebArena and out-of-domain on WebVoyager.


\paragraph{Baselines \& Models.} We compare against NNetNav \citep{murty2024nnetnav}, OS-Genesis \citep{sun2025genesis}, and SynthAgent \citep{wang2025adapting}, which represent self-exploration, random-walk synthesis, and categorized exploration with online refinement, respectively. We fine-tune two open-source multimodal backbones: Qwen3-VL-8B-Instruct \citep{bai2025qwen3} and InternVL3-8B \citep{zhu2025internvl3}. We use Gemini-3-Flash as the base teacher $M_1$ and Gemini-3.1-Pro as the stronger teacher $M_2$.

\paragraph{Data Synthesis Statistics.} Rather than relying on random exploration, SynWeaver first constructs a structured website map for each WebArena website. Table~\ref{tab:webarena_map_stats} reports the retained states and transitions used for synthesis, totaling 559 states and 794 transitions. From this topology, we synthesize 3,500 UI samples and 822 validated task-trajectory pairs comprising 4,500 action steps. For fair comparison, each baseline is trained on 1,000 public trajectories, so SynWeaver uses a smaller but more structured and validated training set.

\begin{table}[t]
  \centering
  \footnotesize
  \setlength{\tabcolsep}{4pt}
  \begin{tabular}{lccccc}
    \toprule[1pt]
    Statistics & Shopping & CMS & Reddit & GitLab & Maps \\
    \midrule[0.6pt]
    States & 90 & 169 & 131 & 153 & 16 \\
    Transitions & 148 & 213 & 203 & 214 & 16 \\
    \bottomrule[1pt]
  \end{tabular}
  \caption{Website map statistics retained for SynWeaver data synthesis on WebArena.}
  \label{tab:webarena_map_stats}
\end{table}

\paragraph{Training Details.} All methods are trained under the same two-stage LoRA pipeline for fair comparison. We first adapt each backbone with SynWeaver's UI data to obtain a shared UI-aware initialization $M_{ui}$, and then fine-tune $M_{ui}$ using the synthesized supervision of each method. Detailed benchmark filtering, environment setup, data sampling rules, and hyperparameters are provided in Appendix~\ref{sec:exp}.

\subsection{Main Results}

\begin{table*}[t]
  \centering
  \footnotesize
  \setlength{\tabcolsep}{6pt}
  \begin{tabular}{clcccccc}
    \toprule[1pt]
    Model & Method & Shopping & CMS & Reddit & GitLab & Maps & Overall \\
    \midrule[0.6pt]
    \multirow[c]{5}{*}{Qwen3-VL-8B-Instruct} & Vanilla & 18.18 & 12.28 & 7.69 & 3.57 & 18.75 & 11.95 \\
     & NNetNav & 20.00 & 14.04 & 11.54 & 5.36 & 21.88 & 14.16 \\
     & OS-Genesis & 20.00 & 12.28 & 11.54 & 10.71 & 28.13 & 15.93 \\
     & SynthAgent & 18.18 & \textbf{15.79} & 15.38 & 5.36 & \textbf{37.50} & 16.81 \\
     & \cellcolor{ourslavender} Ours & \cellcolor{ourslavender}\textbf{23.64} & \cellcolor{ourslavender}14.04 & \cellcolor{ourslavender}\textbf{23.08} & \cellcolor{ourslavender}\textbf{12.50} & \cellcolor{ourslavender}34.38 & \cellcolor{ourslavender}\textbf{19.91} \\
    \midrule[0.6pt]
    \multirow[c]{5}{*}{InternVL3-8B} & Vanilla & 9.09 & 10.53 & 7.69 & 5.36 & 12.50 & 8.85 \\
     & NNetNav & 16.36 & 10.53 & 0.00 & 3.57 & 21.88 & 10.62 \\
     & OS-Genesis & 14.55 & 10.53 & 11.54 & 5.36 & 15.63 & 11.06 \\
     & SynthAgent & 16.36 & 7.02 & \textbf{15.38} & 5.36 & \textbf{28.13} & 12.83 \\
     & \cellcolor{ourslavender} Ours & \cellcolor{ourslavender}\textbf{20.00} & \cellcolor{ourslavender}\textbf{12.28} & \cellcolor{ourslavender}\textbf{15.38} & \cellcolor{ourslavender}\textbf{8.93} & \cellcolor{ourslavender}15.63 & \cellcolor{ourslavender}\textbf{14.16} \\
    \bottomrule[1pt]
  \end{tabular}
  \caption{Task success rate (SR, \%) on WebArena. Bold numbers indicate the best result.}
  \label{tab:main_results_webarena}
\end{table*}

\begin{table*}[t]
  \centering
  \footnotesize
  \setlength{\tabcolsep}{4pt}
  \resizebox{\textwidth}{!}{
  \begin{tabular}{lcccccccccc}
    \toprule[1pt]
    Method & Allrecipes & Amazon & Apple & ArXiv & BBC & ESPN & GitHub & Hugging Face & WA & Overall \\
    \midrule[0.6pt]
    Qwen & 4.44 & 4.88 & 20.93 & 11.63 & 19.05 & 4.55 & 31.71 & 13.95 & 10.87 & 13.40 \\
    +NNetNav & 2.22 & 9.76 & 13.95 & 23.26 & 28.57 & 9.09 & 34.15 & 16.28 & 21.74 & 17.53 \\
    +OS-Genesis & \textbf{6.67} & 21.95 & 27.91 & 18.60 & 21.43 & 11.36 & 29.27 & 23.26 & 15.22 & 19.33 \\
    +SynthAgent & 4.44 & 24.39 & 25.58 & 23.26 & 35.71 & 18.18 & \textbf{36.59} & 20.93 & 15.22 & 22.42 \\
    \cellcolor{ourslavender}+Ours & \cellcolor{ourslavender}2.22 & \cellcolor{ourslavender}\textbf{29.27} & \cellcolor{ourslavender}\textbf{39.53} & \cellcolor{ourslavender}\textbf{27.91} & \cellcolor{ourslavender}\textbf{40.48} & \cellcolor{ourslavender}\textbf{20.45} & \cellcolor{ourslavender}29.27 & \cellcolor{ourslavender}\textbf{30.23} & \cellcolor{ourslavender}\textbf{26.09} & \cellcolor{ourslavender}\textbf{27.06} \\
    \bottomrule[1pt]
  \end{tabular}}
  \caption{Qwen3-VL-8B-Instruct task success rate (SR, \%) on WebVoyager. The abbreviations in the header are BBC = BBC News and WA = Wolfram Alpha. Bold numbers denote the best result in each column.}
  \label{tab:main_results_webvoyager}
\end{table*}

\paragraph{In-domain Results on WebArena.}
As shown in Table~\ref{tab:main_results_webarena}, SynWeaver achieves the best overall SR on both model backbones, despite using only 822 validated task-trajectory pairs after filtering, versus the 1,000 trajectories used for each baseline. On Qwen3-VL-8B-Instruct, our method reaches 19.91, outperforming the strongest baseline, SynthAgent, by 3.10 points and the vanilla model by 7.96 points. On InternVL3-8B, it attains 14.16, improving over the strongest baseline, SynthAgent, by 1.33 points and over the vanilla model by 5.31 points. These gains are noteworthy because, under our matched training setup, all methods are fine-tuned from the same UI-aware initialization $M_{ui}$. This indicates that the advantage mainly comes from the quality of the synthesized task-trajectory pairs rather than from differences in model initialization.

Performance varies across websites and model backbones. For Qwen3-VL-8B-Instruct, our method achieves the best SR on Shopping, Reddit, and GitLab, while SynthAgent remains strongest on CMS and Maps. For InternVL3-8B, our method performs best on Shopping, CMS, and GitLab and ties with SynthAgent for the best result on Reddit. A notable exception is Maps, where our method underperforms SynthAgent across both backbones. A likely reason is that Maps contributes only 48 transition instances for data synthesis, which is substantially fewer than the other websites and therefore provides a much smaller amount of supervision.

\paragraph{Out-of-domain Results on WebVoyager.}
Table~\ref{tab:main_results_webvoyager} shows that SynWeaver also generalizes well to unseen websites. Using Qwen3-VL-8B-Instruct, our method achieves an overall SR of 27.06 on the evaluated WebVoyager websites, outperforming SynthAgent by 4.64 points and the vanilla model by 13.66 points. It attains the best SR on seven of the nine websites, namely Amazon, Apple, ArXiv, BBC, ESPN, Hugging Face, and WA. However, it does not outperform the strongest baseline on Allrecipes and GitHub, indicating that cross-domain transfer remains uneven on these websites.

Overall, the results on both benchmarks support our central hypothesis that high-quality web agent data should couple task intent with executable trajectories. Compared with baselines that rely on self-exploration, random walks, or decoupled task-trajectory refinement, SynWeaver consistently delivers stronger overall performance.  Notably, this improvement is achieved with 822 validated task-trajectory pairs after filtering, which is fewer than the 1,000 trajectories used for each baseline. These findings suggest that our gains come not only from better alignment, but also from better data efficiency.
\section{Analysis}

\subsection{Ablation Study}

\begin{table}[t]
  \centering
  \footnotesize
  \setlength{\tabcolsep}{4pt}
  \resizebox{\columnwidth}{!}{%
  \begin{tabular}{lcccccc}
    \toprule[1pt]
    Method & Shopping & CMS & Reddit & GitLab & Maps & Overall \\
    \midrule[0.6pt]
    Ours & 23.64 & \textbf{14.04} & \textbf{23.08} & \textbf{12.50} & \textbf{34.38} & \textbf{19.91} \\
    $-$Map & 20.00 & 8.77 & 15.38 & 8.93 & 31.25 & 15.49 \\
    $-$CR & 18.18 & 10.53 & 15.38 & 5.36 & \textbf{34.38} & 15.04 \\
    $-$WP & \textbf{27.27} & 12.28 & 19.23 & 8.93 & 25.00 & 17.70 \\
    $-$PV & 21.82 & 7.02 & 19.23 & 7.14 & 25.00 & 14.60 \\
    \midrule[0.6pt]
    Qwen & 18.18 & 12.28 & 7.69 & 3.57 & 18.75 & 11.95 \\
    +UI & 16.36 & 10.53 & 11.54 & 5.36 & \textbf{34.38} & 14.16 \\
    +Traj & 25.45 & \textbf{14.04} & 19.23 & 8.93 & 31.25 & 18.58 \\
    \bottomrule[1pt]
  \end{tabular}}
  \caption{Ablation results (SR, \%) on WebArena using Qwen3-VL-8B-Instruct. Ours denotes the full SynWeaver pipeline. $-$Map replaces website map with random walk; $-$CR replaces collaborative refinement with decoupled refinement; $-$WP replaces the website-prior proposer with a general-purpose model; and $-$PV removes post-verification. UI and Traj use only UI data and trajectory data, respectively, for training.}
  \label{tab:ablation_webarena}
\end{table}

We conduct an ablation study on Qwen3-VL-8B-Instruct over the WebArena subset, as shown in Table~\ref{tab:ablation_webarena}. The upper block ablates components of the full SynWeaver pipeline, while the lower block compares the Qwen baseline with training on UI data only (+UI) or trajectory data only (+Traj). Relative to Ours, replacing website map  sampling with random walk ($-$Map) reduces the overall SR by 4.42 points, demonstrating the value of structured website exploration. Replacing collaborative refinement with decoupled refinement ($-$CR) causes a 4.87-point drop, with the largest losses on Reddit (7.70) and GitLab (7.14), confirming the benefit of jointly refining tasks and trajectories. Replacing the website-prior proposer with Gemini-3.1-Pro ($-$WP) lowers the overall SR by 2.21 points and particularly hurts Maps (9.38), while removing post-verification ($-$PV) produces the largest overall drop of 5.31 points. In the lower block, +UI and +Traj improve the overall SR over Qwen by 2.21 and 6.63 points, respectively. Overall, the full pipeline performs best.

\subsection{Impact of Website Maps on Data Diversity}

To verify whether tasks synthesized from website maps are more diverse, we compare the synthesized task sets in Table~\ref{tab:diversity_analysis}. We use NovelSum \citep{yang2025measuring}, which summarizes set-level novelty using both semantic difference and local density; here, $K$, $\alpha$, and $\beta$ control the neighborhood size, distance weighting, and density effect, respectively. For evaluation, we use the 1,009 human-annotated training tasks from Mind2Web \citep{deng2023mind2web} as the reference set, encode task texts with BGE-M3 \citep{chen2024m3}, and report three settings to ensure robust results: the standard setting $(K{=}10, \beta{=}0.5, \alpha{=}1)$, the density-free variant $(K{=}10, \beta{=}0, \alpha{=}1)$, and the variant with different neighbors $(K{=}5, \beta{=}0.5, \alpha{=}1)$. We also report cosine distance as a complementary measure. SynWeaver achieves the best NovelSum in all three settings (0.3805, 0.3731, and 0.3854), while remaining essentially tied with the strongest baseline on cosine distance. This shows that website maps improve task diversity without sacrificing overall semantic spread.

\begin{table}[t]
  \centering
  \footnotesize
  \setlength{\tabcolsep}{4pt}
  \resizebox{\columnwidth}{!}{%
  \begin{tabular}{lcccc}
    \toprule[1pt]
    \multirow{2}{*}{Method} & \multicolumn{3}{c}{NovelSum} & \multirow{2}{*}{Cosine Distance} \\
    \cmidrule(lr){2-4}
     & Setting 1 & Setting 2 & Setting 3 & \\
    \midrule[0.6pt]
    NNetNav & 0.3585 & 0.3359 & 0.3640 & 0.5517 \\
    OS-Genesis & 0.3624 & 0.3528 & 0.3673 & 0.5606 \\
    SynthAgent & 0.3725 & 0.3681 & 0.3770 & \textbf{0.5825} \\
    Ours & \textbf{0.3805} & \textbf{0.3731} & \textbf{0.3854} & 0.5820 \\
    \bottomrule[1pt]
  \end{tabular}}
  \caption{Data diversity comparison across different synthesis methods. Settings 1, 2, and 3 correspond to $(K{=}10, \beta{=}0.5, \alpha{=}1)$, $(K{=}10, \beta{=}0, \alpha{=}1)$, and $(K{=}5, \beta{=}0.5, \alpha{=}1)$, respectively. Higher values indicate greater diversity.}
  \label{tab:diversity_analysis}
\end{table}

\subsection{Impact of Website-Prior on Task Quality}

To assess the effect of website-prior on task proposal, we compare the UI-aware model with Gemini-3.1-Pro on the same transition triplets as Table~\ref{tab:webarena_map_stats}, changing only the proposer. Tasks from both models undergo the same Collaborative Trajectory Synthesis procedure. We report final task success rate and the average number of task refinement, where higher success and fewer refinements indicate better-grounded proposals.

\begin{figure}[t]
  \centering
  \includegraphics[width=\columnwidth]{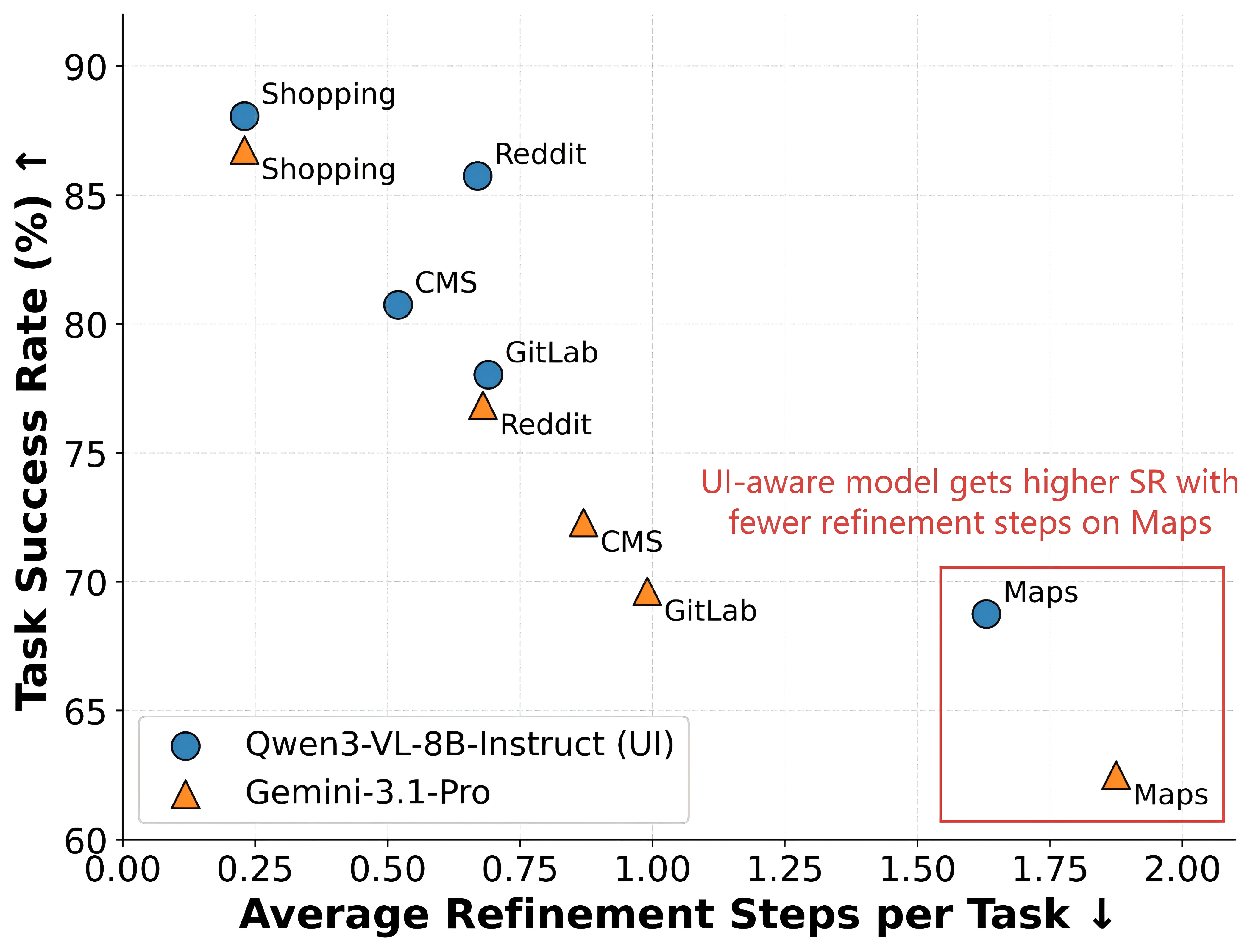}
  \caption{Effect of website-prior on task proposal quality across WebArena websites.}
  \label{fig:wp_result}
\end{figure}

Figure~\ref{fig:wp_result} shows that the UI-aware proposer achieves higher task success rates with fewer optimizations across all WebArena websites. We do not control for task difficulty, so simpler tasks may naturally score better on both metrics. Nevertheless, Table~\ref{tab:ablation_webarena} shows that replacing the website-prior proposer with a general-purpose model ($-$WP) reduces downstream SR from 19.91 to 17.70, indicating that website-prior proposals and their trajectories provide more effective training supervision.

\subsection{Impact of Collaborative Refinement on Data Efficiency and Cost}

To examine how collaborative refinement affects data efficiency and synthesis cost, we compare it with decoupled refinement on the same task set. Decoupled refinement updates tasks during trajectory collection and revises trajectories after execution, whereas collaborative refinement jointly updates the task specification and executed prefix when inconsistencies arise. We report task success rate, trajectory retention, and estimated synthesis cost, with SynthAgent included as a cost reference.

\begin{table}[t]
  \centering
  \footnotesize
  \setlength{\tabcolsep}{3.5pt}
  \resizebox{\columnwidth}{!}{%
  \begin{tabular}{llrrr}
    \toprule[1pt]
    Method & Teacher Model & \shortstack[c]{Task SR (\%)} & \shortstack[c]{Retention (\%)} & Cost (\$) \\
    \midrule[0.5pt]
    SynthAgent & GPT-4.1 & -- & -- & 0.130 \\
    \midrule[0.5pt]
    DR & \multirow{2}{*}{\shortstack[l]{Gemini-3-Flash\\+ Gemini-3.1-Pro}} & 71.89 & 90.59 & 0.091 \\
    CR &  & \textbf{74.79} & \textbf{99.52} & \textbf{0.085} \\
    \bottomrule[1pt]
  \end{tabular}}
  \caption{Comparison between decoupled (DR) and collaborative (CR) refinement on the same task set. SynthAgent is included as a cost reference.}
  \label{tab:collab_efficiency_cost}
\end{table}

As shown in Table~\ref{tab:collab_efficiency_cost}, collaborative refinement improves task success rate from 71.89\% to 74.79\% and raises trajectory retention from 90.59\% to 99.52\%. The main gain comes from retention: far fewer trajectories are discarded after verification, so more collected data can be retained as supervision. This higher retention directly lowers synthesis cost, reducing it from \$0.091 to \$0.085 despite better task success and data quality. Cost is further reduced by the hierarchical teacher design, where the lightweight teacher $M_1$ handles routine synthesis and the stronger teacher $M_2$ is used only for difficult refinements and repairs. As a result, SynWeaver is also cheaper than SynthAgent (\$0.13). Overall, collaborative refinement improves both data efficiency and economic efficiency.

\section{Conclusion}

In this paper, we present SynWeaver, a website-prior task-trajectory co-synthesis framework for web agents. SynWeaver addresses a key limitation of existing exploration-based synthesis methods: they often lack sufficient website-specific priors to propose realistic and executable tasks, leading to limited trajectory quality. To overcome this issue, SynWeaver combines exploration-driven website map construction with website-prior learning and collaborative task-trajectory refinement, producing supervision that is more faithfully aligned with real website interactions. Experiments on WebArena and WebVoyager demonstrate that SynWeaver consistently outperforms prior synthesis baselines while also improving data diversity and synthesis efficiency. These results highlight the importance of explicit website knowledge in scalable web agent data synthesis and provide a practical direction for improving web agent adaptation and generalization.

\section*{Limitations}

SynWeaver has limitations in both deployment scope and training paradigm. Its website maps are constructed through automatic exploration; although this process is effective in self-hostable and moderately constrained environments, it becomes less reliable when websites adopt strong anti-bot or human-verification mechanisms, such as CAPTCHA systems, aggressive rate limiting, or dynamic authentication pipelines. In addition, SynWeaver mainly improves web agents through supervised fine-tuning on teacher-generated data, making the resulting student model inherently bounded by the reasoning and interaction capabilities of the proprietary teacher models used during synthesis. The current framework also does not explore more autonomous self-improving paradigms, such as online reinforcement learning, which may further enhance agents beyond the supervision provided by the teachers. We leave these directions to future work.

\section*{Ethical Considerations}

All experiments in this work are conducted in controlled, compliant, and authorized web environments, such as benchmarks and self-hosted websites. We do not crawl third-party websites without permission, bypass access controls, collect private user data, or disrupt online services. Because automated website exploration could be misused for unauthorized crawling, large-scale scraping, or other harmful activities, we strongly oppose such uses and call on researchers and practitioners to obtain proper authorization, respect website policies and rate limits, and use these methods only for lawful and responsible purposes.

\section*{Use of AI Assistants}

AI assistants were used only for grammar checking and language polishing during manuscript preparation. All scientific ideas, methods, experiments, results, analyses, and conclusions were developed and verified by the authors. The authors manually reviewed and edited all AI-assisted text and take full responsibility for the final manuscript.



\bibliography{custom}

\newpage

\appendix

\section{Website Crawling Details}
\label{sec:crawler}

This appendix describes the construction of the website map $\mathcal{G}$. Starting from a seed URL, the crawler incrementally explores reachable pages with depth-first search (DFS) and records both states and transitions in a persistent graph. Algorithm~\ref{alg:main_crawl} is the main controller: it acquires the current state, inserts the state and transition into $\mathcal{G}$, filters candidate actions, and recursively explores newly discovered states. Algorithm~\ref{alg:dedup} is called inside this DFS loop to prune redundant lists, forms, and atomic elements before execution. Algorithm~\ref{alg:state_equivalent} is applied after each interaction to determine whether the resulting page should be treated as a new state. Algorithm~\ref{alg:backtrack} is executed after recursive exploration returns, so that the crawler can recover the previous action context and continue exploring the remaining actions from the correct state. Table~\ref{tab:symbols} summarizes the notation shared by these algorithms, and Table~\ref{tab:functions} defines the principal graph and browser operations used in their pseudocode.

\paragraph{Why we adopt DFS?} We adopt DFS as the outer exploration strategy because it is substantially more efficient than breadth-first search (BFS) in an interactive browser environment. Under BFS, exploration proceeds level by level. After expanding one state, the crawler must repeatedly return to earlier states and then restore the browser to another frontier state in order to continue expansion. In practice, this requires frequent rollback and replay operations, such as navigating back, reopening pages, or deterministically replaying prior actions, which introduces considerable browser overhead. By contrast, DFS continues along the current branch and performs recovery mainly when the branch has been exhausted, thereby greatly reducing the number of restoration operations and the associated resource cost.

\paragraph{State.} A page state $s_d=(u, E, L, F, a11y, img)$ is a structured snapshot of the current web interface. Specifically, $u$ is the current URL, $E$ is the set of actionable elements detected on the page, $L$ is the set of extracted list containers, $F$ is the set of extracted form containers, and $(a11y, img)$ denote the accessibility tree and screenshot, respectively. Here, a list container is detected from DOM containers such as \texttt{ul}, \texttt{ol}, \texttt{table}/\texttt{tbody}, or list-like ARIA containers (e.g., \texttt{role="list"}, \texttt{role="listbox"}, and \texttt{role="menu"}). A container is treated as a valid list only if it contains over three actionable list items overall. A form container is detected only from a genuine ancestor \texttt{<form>} element. Elements are grouped by their nearest form ancestor, and a container is retained as a valid form only if it contains at least one input field and one submit button.

\paragraph{Transition.} A transition is defined as $\tau_d=(s_d, s_{d+1}, newPageOpened, op_d)$, where $s_d$ is the pre-action state, $s_{d+1}$ is the resulting state after execution, $newPageOpened$ indicates whether the action creates a new page, and $op_d$ records the executed operation. This representation makes each edge in the website graph self-contained: it captures not only how the crawler moves from one state to another, but also the information needed for later recovery and replay. In practice, transitions are produced by three kinds of execution routines: form execution, atomic element execution, and page-level scrolling. Among them, atomic execution is rule-based: the crawler chooses an action according to the target element's tag, type, text, and accessibility attributes. Table~\ref{tab:action_tag_mapping} summarizes the action categories used in our implementation, together with their high-level semantics and the tag patterns to which they apply. If the element does not match any action, \texttt{click} is executed by default.

\begin{table*}[!t]
  \centering
  \renewcommand{\arraystretch}{1.12}
  \begin{tabular}{l p{0.25\linewidth} p{0.50\linewidth}}
    \hline
    \textbf{Action} & \textbf{Description} & \textbf{Applicable Tags / Targets} \\
    \hline
    \texttt{click} &
    Clicks the element and then checks whether the interaction opens a new tab. &
    Direct click targets: \texttt{<button>}, \texttt{<a>}, \texttt{<summary>}, \texttt{<details>}, \texttt{<img>}, \texttt{<span>}, \texttt{<video>}, \texttt{<audio>}, \texttt{<iframe>}, \texttt{<label>}; \texttt{<input>} with clickable types (\texttt{checkbox}, \texttt{radio}, \texttt{button}, \texttt{submit}, \texttt{reset}, \texttt{image}) \\
    \hline
    \texttt{fill} &
    Fills an input control with a default value based on the field type and textual attributes. &
    \texttt{<textarea>}; \texttt{<input>} with input types (\texttt{text}, \texttt{email}, \texttt{password}, \texttt{tel}, \texttt{url}, \texttt{number}, and \texttt{date}), excluding search-like fields \\
    \hline
    \texttt{select} &
    Selects an option from a dropdown menu. &
    \texttt{<select>} \\
    \hline
    \texttt{hover} &
    Hovers over the target element to reveal hidden menus. &
    \texttt{<li>} with submenu cues in text or \texttt{aria-label}; fallback elements whose \texttt{aria-label} indicates menu or dropdown behavior \\
    \hline
    \texttt{scroll} &
    Scrolls the page by one viewport. &
    N/A \\
    \hline
    \texttt{form\_submission} &
    Executes a composite form action: fills fields and then submits form. &
    Extracted \texttt{<form>} containers, typically with \texttt{<input>}, \texttt{<textarea>}, or \texttt{<select>} fields and a submit control such as \texttt{<button>} or \texttt{<input type="submit">} \\
    \hline
  \end{tabular}
  \caption{Browser actions for website map construction and their applicable tag patterns.}
  \label{tab:action_tag_mapping}
\end{table*}

\paragraph{Duplicate-trigger Detection.} Duplicate-trigger detection is required to ensure that the website graph covers each distinct functional state while avoiding repeated exploration of equivalent functionality. To support this goal, the crawler maintains an exploration registry $\mathcal{M}=(V_e, V_l, X_l, X_f, D_u)$ across recursive calls. The set $V_e$ records element signatures so that once an atomic element with the same functional profile has been triggered, the crawler does not need to trigger it again on later pages. However, element signatures alone are not sufficient, because two elements may serve the same function while carrying different local text. A typical example is a content feed such as Reddit. There are many posts sharing the same structure and affording the same interactions, but their texts differ, so they may still be treated as different elements by signature matching. To address this case, $D_u$ records visited element XPaths under the same URL template, allowing the crawler to deduplicate elements that appear at the same structural position on template-equivalent pages. The same design principle is used for lists. The set $V_l$ stores list signatures to suppress obviously identical list containers, but lists on structurally similar pages may still provide the same functionality even when their item texts or item counts differ. Therefore, $X_l$ additionally records visited list-container XPaths so that functionally equivalent lists are not re-explored solely because their contents change. For forms, repeated functionality is less likely to reappear across different page structures, so the crawler adopts a lighter strategy and records only XPaths of form containers in $X_f$, which saves memory while remaining sufficient in practice. In Algorithm~\ref{alg:dedup}, signatures are computed by $\texttt{GetSign}$ as lightweight identifiers rather than full semantic hashes. For an atomic element, the signature is formed by concatenating its tag, name, aria-label, and text, i.e., $\textit{sig}(e)=\textit{tag}\mid\textit{name}\mid\textit{aria\_label}\mid\textit{text}$. For a list container, the signature is computed by concatenating the container tag with the texts of its list items, equivalently the container tag plus the string obtained by prefixing each item text with a separator and concatenating them in order. This design keeps signature computation simple and efficient, while the combination of signature and XPath prevents redundant exploration more robustly than any single mechanism alone.

\paragraph{Constrained Exploration.} To ensure finite and effective website exploration, the crawler relies not only on the duplicate-trigger detection mechanism described above, but also on three additional constraints applied during each recursive step: $\texttt{ShouldStop}$, $\texttt{IsExitGuarded}$, and $\texttt{SampleItems}$. First, $\texttt{ShouldStop}$ determines whether the current branch should terminate immediately. In practice, this happens when the crawler fails to obtain the current state, when the graph $\mathcal{G}$ detects that the state is already known, when the current URL leaves the target root domain and enters an external domain, when the current URL matches a manually configured blocked URL, or when the predefined exploration budget, such as the maximum depth or the maximum number of states, has been reached. Second, $\texttt{IsExitGuarded}$ checks whether a candidate element belongs to a predefined set of forbidden interactions before the crawler executes it. This mechanism is mainly used to block elements that may prematurely terminate or disrupt the session, such as logout, sign out, or account deletion controls, and thus prevents the crawler from leaving the intended exploration workflow through unsafe actions. Third, $\texttt{SampleItems}$ constrains exploration within list containers by selecting only representative elements rather than exhaustively triggering every repeated item. The key observation is that many lists contain multiple items with the same functional structure, so exploring all of them would add cost without revealing new behavior. In our framework, sampling is primarily guided by an LLM: if the list items are judged to provide different functions, all of them are retained; if they are judged to contain repeated functional groups, only one representative group is kept. For example, in a product-card list, each card may contain the same interaction pattern, such as product image, product name, add-to-cart, and add-to-wishlist, while differing only in product content, in which case exploring a single representative card is sufficient.

\paragraph{Backtrack.} After exploring the subtree induced by one action, the crawler must restore the pre-action context before it can continue with the remaining unexplored actions. Algorithm~\ref{alg:backtrack} therefore uses three recovery strategies in order. The first strategy closes a newly opened tab, which directly handles actions that spawn a separate browsing page. The second strategy uses browser history navigation when the interaction changes the current page within the same tab. The third strategy replays the shortest path from the root state, which is more expensive but more robust when local rollback is unavailable or unreliable. This prioritized design keeps common recovery cases lightweight while still preserving a fallback for dynamic or non-reversible interactions.

\begin{table}[!t]
  \centering
  \begin{tabular}{l p{0.7\linewidth}}
    \hline
    \textbf{Symbol} & \textbf{Description} \\
    \hline
    $\Gamma$ & Exploration configuration \\
    $\mathcal{B}$ & Browser interaction agent that executes browser operations \\
    $\mathcal{G}$ & Website graph \\
    $\mathcal{M}$ & Exploration registry, $\mathcal{M}=(V_e, V_l, X_l, X_f, D_u)$ \\
    $d$ & Current exploration depth \\
    $s_d$ & State at depth $d$, where $s_d = (u, E, L, F, a11y, img)$ \\
    $\tau_d$ & Transition generated at depth $d$, where $\tau_d = (s_d, s_{d+1}, newPageOpened, op_d)$ \\
    $\omega_l \in L$ & List container in the current state \\
    $\omega_f \in \Omega_f$ & Form in the current state \\
    $\omega_e \in \Omega_e$ & Atomic element in the current state \\
    $u$ & URL \\
    $E$ & Set of all actionable elements on the webpage \\
    $L$ & Set of all list containers on the webpage \\
    $F$ & Set of all form containers on the webpage \\
    $a11y$ & Accessibility tree \\
    $img$ & Screenshot of the webpage \\
    $V_e$ & Set of visited element signatures \\
    $V_l$ & Set of visited list signatures \\
    $X_l$ & Set of visited list XPaths \\
    $X_f$ & Set of visited form XPaths \\
    $D_u$ & Map from a URL template to visited element XPaths \\
    \hline
  \end{tabular}
  \caption{Notation Description.}
  \label{tab:symbols}
\end{table}

\begin{table}[!t]
  \centering
  \begin{tabular}{l p{0.5\linewidth}}
    \hline
    \textbf{Function} & \textbf{Description} \\
    \hline
    $\texttt{Navigate}$ & Navigates to the specified URL $u$. \\
    $\texttt{GetCurrentState}$ & Returns the current webpage state $s_d$. \\
    $\texttt{AddState}$ & Adds the state $s_d$ to the graph $\mathcal{G}$ and returns whether it is a duplicate $dup$. \\
    $\texttt{AddTransition}$ & Adds the transition $\tau_d$ to the graph $\mathcal{G}$. \\
    $\texttt{ShouldStop}$ & Determines whether current exploration branch should be stopped. \\
    $\texttt{IsExitGuarded}$ & Checks whether the element $\omega_e$ is in the block list. \\
    $\texttt{ExecuteForm}$  & Replays a recorded form operation when available; otherwise, it fills and submits the form with the LLM, and falls back to default values if needed. \\
    $\texttt{ExecuteAtomic}$ & Replays a recorded operation when available; Otherwise, it triggers the element according to its tag. \\
    $\texttt{GetSign}$ & Computes the signature of the object. \\
    $\texttt{GetItems}$ & Returns all elements contained in the given container. \\
    $\texttt{GetShortestPath}$ & Computes the shortest path from the root state to the target state using bidirectional breadth-first search. \\
    \hline
  \end{tabular}
  \caption{Function Description.}
  \label{tab:functions}
\end{table}

\begin{algorithm}[t]
\caption{DFS-based Website Crawling}
\label{alg:main_crawl}
\begin{algorithmic}[1]
\Require{Seed URL $u_0$, configuration $\Gamma$}
\Ensure{Persistent graph $\mathcal{G}$}
\State Initialize $\mathcal{B} \gets \texttt{InitAgent}(\Gamma)$
\State Initialize $\mathcal{G} \gets \texttt{InitGraph}(\Gamma)$
\State Initialize $\mathcal{M} \gets \texttt{InitRegistry}()$
\State $\mathcal{B}.\texttt{Navigate}(u_0)$
\State $\texttt{Explore}(\perp, \perp, 0, \mathcal{B}, \mathcal{G}, \mathcal{M}, \Gamma)$
\State \Return{$\mathcal{G}$}

\vspace{0.4em}
\Procedure{Explore}{$s_{d-1}$, $\tau_{d-1}$, $d$, $\mathcal{B}$, $\mathcal{G}$, $\mathcal{M}$, $\Gamma$}
    \State $s_{d} \gets \mathcal{B}.\texttt{GetCurrentState}()$
    \State $(u_d, \_, \_, \_, \_, \_) \gets s_d$
    \State $dup \gets \mathcal{G}.\texttt{AddState}(s_{d})$
    \Statex \hspace{\algorithmicindent} where $dup$ indicates whether $s_d$ has already been visited.
    \If{$s_{d-1} \neq \perp \land \tau_{d-1} \neq \perp$}
        \State $ \mathcal{G}.\texttt{AddTransition}(\tau_{d-1}) $
    \EndIf
    \If{$\texttt{ShouldStop}(dup, s_{d}, d, \mathcal{G}, \Gamma)$}
        \State \Return{}
    \EndIf
    \State $ (\Omega_f, \Omega_e) \gets \texttt{FilterState}(s_{d}, \mathcal{M}, \Gamma)$
    \For{$\omega_f \in \Omega_f$}
        \State $(s_{d+1}, \tau_{d}) \gets \mathcal{B}.\texttt{ExecuteForm}(\omega_f, s_{d})$
        \State $eq \gets \texttt{StateEquiv}(s_{d}, s_{d+1}, \Gamma)$
        \If{$\neg eq$}
            \State $\texttt{Explore}(s_{d}, \tau_{d}, d+1, \mathcal{B}, \mathcal{G}, \mathcal{M}, \Gamma)$
            \State $\texttt{Backtrack}(\tau_{d}, \mathcal{B}, \mathcal{G}, \Gamma)$
        \EndIf
    \EndFor
    \For{$\omega_e \in \Omega_e$}
        \If{$\texttt{IsExitGuarded}(\omega_e, \Gamma)$}
            \State \textbf{continue}
        \EndIf
        \State $(s_{d+1}, \tau_{d}) \gets \mathcal{B}.\texttt{ExecuteAtomic}(\omega_e, s_{d})$
        \State $eq \gets \texttt{StateEquiv}(s_{d}, s_{d+1}, \Gamma)$
        \If{$\neg eq$}
            \State $\texttt{Explore}(s_{d}, \tau_{d}, d+1, \mathcal{B}, \mathcal{G}, \mathcal{M}, \Gamma)$
            \State $\texttt{Backtrack}(\tau_{d}, \mathcal{B}, \mathcal{G}, \Gamma)$
        \EndIf
    \EndFor
    \State $(s_{d+1}, \tau_{d}) \gets \mathcal{B}.\texttt{ExecuteScroll}(s_{d})$
    \Comment{Scroll the page}
    \State $eq \gets \texttt{StateEquiv}(s_{d}, s_{d+1}, \Gamma)$
    \If{$\neg eq$}
        \State $\texttt{Explore}(s_{d}, \tau_{d}, d+1, \mathcal{B}, \mathcal{G}, \mathcal{M}, \Gamma)$
        \State $\mathcal{B}.\texttt{Navigate}(u_d)$
    \EndIf
\EndProcedure
\end{algorithmic}
\end{algorithm}

\begin{algorithm}[t]
\caption{Duplicate-trigger Detection}
\label{alg:dedup}
\begin{algorithmic}[1]
\Require{
Current state $s_{d}$, registry 
$\mathcal{M}$, configuration $\Gamma$
}
\Ensure{Forms $\Omega_f$, raw elements $\Omega_e$}
\Procedure{FilterState}{$s_{d},\mathcal{M},\Gamma$}
    \State $\Omega_f \gets \varnothing,\;\Omega_e \gets \varnothing,\;\Delta \gets \varnothing$
    \Statex \hspace{\algorithmicindent} where $\Delta$ stores elements pruned from further exploration.
    \State $(u_d, E, L, F, \_, \_) \gets s_d$
    \State $(V_e, V_l, X_l, X_f, D_u) \gets \mathcal{M}$
    \State $t \gets \texttt{GetTemplate}(u_d)$
    \Comment{Get the URL template}

    \vspace{0.3em}
    \Statex \textit{// List-level pruning}
    \For{$\omega_l \in L$}
        \If{$\texttt{GetSign}(\omega_l) \in V_l \;\lor\; \omega_l.xpath\in X_l$}
            \State $\Delta \gets \Delta \cup \texttt{GetItems}(\omega_l)$
        \Else
            \State $I_{\omega_l} \gets \texttt{SampleItems}(\omega_l, \Gamma)$
            \Comment{Sample representative items}
            \State $\Delta \gets \Delta \cup (\texttt{GetItems}(\omega_l)\setminus I_{\omega_l})$
        \EndIf
        \State $V_l \gets V_l \cup \{\texttt{GetSign}(\omega_l)\}$
        \State $X_l \gets X_l \cup \{\omega_l.xpath\}$
    \EndFor

    \vspace{0.3em}
    \Statex \textit{// Form-level pruning}
    \For{$f \in F$}
        \State $E_f \gets \texttt{GetItems}(f)$
        \If{$f.xpath \notin X_f$}
            \State $\Omega_f \gets \Omega_f \cup \{f\}$
        \EndIf
        \State $\Delta \gets \Delta \cup E_f$
        \State $X_f \gets X_f \cup \{f.xpath\}$
    \EndFor

    \vspace{0.3em}
    \Statex \textit{// Element-level deduplication}
    \For{$e \in E$}
        \If{$e \in \Delta \; \lor\; \texttt{GetSign}(e) \in V_e \;\lor\; e.xpath \in D_u[t]$}
            \State \textbf{continue}
        \EndIf
        \State $\Omega_e \gets \Omega_e \cup \{e\}$
        \State $V_e \gets V_e \cup \{\texttt{GetSign}(e)\}$
        \State $D_u[t] \gets D_u[t] \cup \{e.xpath\}$
    \EndFor
    \State \Return $(\Omega_f,\Omega_e)$
\EndProcedure
\end{algorithmic}
\end{algorithm}

\begin{algorithm}[t]
\caption{Progressive State Comparison}
\label{alg:state_equivalent}
\begin{algorithmic}[1]
\Require{Current state $s_{d}$, candidate state $s_{d+1}$, configuration $\Gamma$}
\Ensure{$eq$}

\Procedure{StateEquiv}{$s_{d}, s_{d+1}, \Gamma$}
    \State $(u_d, \_, \_, \_, a11y_d, img_d) \gets s_d$
    \State $(u_{d+1}, \_, \_, \_, a11y_{d+1}, img_{d+1}) \gets s_{d+1}$
    \State $(a11yTolerance, useVLM) \gets \Gamma$  
    
    \Statex \textit{// Level 1: URL consistency}
    \If{$u_d \neq u_{d+1}$}
        \State \Return $\textbf{false}$
    \EndIf

    \Statex \textit{// Level 2: Accessibility-tree similarity}
    \State $C_d \gets [\;]$, $C_{d+1} \gets [\;]$
    \ForAll{row $r$ in $a11y_d$}
        \State append the normalized row $r$ to $C_d$
    \EndFor
    \ForAll{row $r$ in $a11y_{d+1}$}
        \State append the normalized row $r$ to $C_{d+1}$
    \EndFor
    \State $N_d \gets |C_d|,\; N_{d+1} \gets |C_{d+1}|,\; N \gets \max(N_d,N_{d+1})$
    \If{$|N_d-N_{d+1}| < N\cdot(1-a11yTolerance)$}
        \State $m \gets 0$
        \For{$i=1$ to $\min(N_d,N_{d+1})$}
            \State remove all digits from $C_d[i]$ and $C_{d+1}[i]$
            \If{the transformed rows are equal}
                \State $m \gets m + 1$
            \EndIf
        \EndFor
        \State $sim \gets m / N$
        \If{$sim \ge a11yTolerance$}
            \State \Return $\textbf{true}$
        \EndIf
    \EndIf

    \Statex \textit{// Level 3: Visual fallback}
    \If{$useVLM$}
        \If{$\texttt{CompareByVLM}(img_d, img_{d+1})=\textbf{true}$}
            \State \Return $\textbf{true}$
        \Else
            \State \Return $\textbf{false}$
        \EndIf
    \EndIf
    \State \Return $\textbf{false}$
\EndProcedure
\end{algorithmic}
\end{algorithm}

\begin{algorithm}[t]
\caption{Backtracking to a Target State}
\label{alg:backtrack}
\begin{algorithmic}[1]
\Require{
Current transition $\tau_d$, agent $\mathcal{B}$, graph $\mathcal{G}$, configuration $\Gamma$
}
\Ensure{Backtrack success flag}
\Procedure{Backtrack}{$\tau_d, \mathcal{B}, \mathcal{G}, \Gamma$}
    \State $(s_d, s_{d+1}, newPageOpened, op_d) \gets \tau_d$
    
    \Statex \textit{// Strategy 1: close newly opened page}
    \If{$newPageOpened = \textbf{true} \land \mathcal{B}.\texttt{BackToPreviousPage}()$}
        \Comment{Close the new page and back to the previous page}
        \State $s' \gets \mathcal{B}.\texttt{GetCurrentState}()$
        \If{$\texttt{StateEquiv}(s_d, s', \Gamma)$}
            \State \Return{$\textbf{true}$}
        \EndIf
    \EndIf

    \Statex \textit{// Strategy 2: browser history back}
    \State $(u_d, \_, \_, \_, \_, \_) \gets s_d$
    \State $(u_{d+1}, \_, \_, \_, \_, \_) \gets s_{d+1}$
    \If{$u_d \neq u_{d+1} \land \mathcal{B}.\texttt{NavigateBack}()$}
        \Comment{Back to the previous page.}
        \State $s' \gets \mathcal{B}.\texttt{GetCurrentState}()$
        \If{$\texttt{StateEquiv}(s_d, s', \Gamma)$}
            \State \Return{$\textbf{true}$}
        \EndIf
    \EndIf

    \Statex \textit{// Strategy 3: deterministic replay from root}
    \State $P \gets \mathcal{G}.\texttt{GetShortestPath}(s_d)$
    \State $u_0 \gets \mathcal{B}.\texttt{GetSeedURL}()$
    \State $\mathcal{B}.\texttt{ResetContext}()$
    \State $\mathcal{B}.\texttt{Navigate}(u_0)$
    \For{$\tau \in P$}
        \State $(\_, \_, \_, op) \gets \tau$
        \If{$op$.\texttt{isForm}}
            \State $\mathcal{B}.\texttt{ExecuteForm}(op)$
        \Else
            \State $\mathcal{B}.\texttt{ExecuteAtomic}(op)$
        \EndIf
    \EndFor

    \State $s' \gets \mathcal{B}.\texttt{GetCurrentState}()$
    \State \Return{$\texttt{StateEquiv}(s_d, s', \Gamma)$}
\EndProcedure
\end{algorithmic}
\end{algorithm}

\section{Heuristic Verification Details}
\label{sec:pos-val}

Heuristic verification is the first stage of post-verification and is applied to a completed candidate trajectory $h^e=(o_1,r_1,a_1,\dots,o_H,r_H,a_H)$. The procedure is purely offline: it does not re-execute the browser or query the environment again, but instead diagnoses trajectory quality from the finished interaction record. For non-empty trajectories, it applies three deterministic checks in order: (1) \textit{termination effectiveness}, (2) \textit{trajectory validity}, and (3) \textit{trajectory consistency}. A candidate passes heuristic verification if and only if no issue is detected.

\paragraph{Termination Effectiveness.} A valid trajectory is required to end with an explicit completion action, denoted by \texttt{none}. This convention provides a canonical marker that the task objective has been achieved and aligns the trajectory with a final completion statement rather than a mere cessation of interaction. Therefore, if the last step is not \texttt{none}, the trajectory is judged to be missing an explicit ending signal and fails termination effectiveness.

\paragraph{Trajectory Validity.} Beyond the explicit ending marker, heuristic verification checks whether the completed trajectory is a plausible successful execution artifact. First, an empty trajectory is rejected immediately, since no executable evidence of task completion is present. Second, let $n=|h^e|$ and let $B$ denote the predefined step budget. When $n \ge B$, the trajectory is flagged with an error. Excessive length often reflects program vulnerabilities because we have set the step limit before execution. Third, the program will detect the state of the environment during execution. If the execution result is labelled as failure, it fails the test. This often happens when some behaviors of the agent cause environmental errors.

\paragraph{Trajectory Consistency.} The consistency check is designed to detect cyclic or stalled behavior. We first map $h^e$ to an action-signature sequence $\Sigma=(\sigma_1,\dots,\sigma_n)$, where each $\sigma_i$ preserves only the information that determines functional behavior: the action type, the identity of the target when applicable, and the action value when the semantics of that action depend on it (e.g. \texttt{type:[4480]:pencil}). This abstraction suppresses incidental observation differences while preserving materially distinct operations. The loop search is activated only when $n \ge 4$. It then enumerates candidate pattern lengths $\ell \in [1,\min(\lfloor n/2 \rfloor,10)]$ and, for each $\ell$, enumerates feasible start positions $j$. For each candidate pair $(j,\ell)$, define
\[
p_{j,\ell} = (\sigma_j,\dots,\sigma_{j+\ell-1}).
\]
Starting from the next block of length $\ell$, the verifier checks whether subsequent blocks match $p_{j,\ell}$ exactly and contiguously. If no such repetition exists, it moves to the next candidate; if repetition is observed, it continues counting until the first mismatch, at which point the current candidate stops expanding. Let $r_{j,\ell}$ denote the number of contiguous repeats observed after the initial block. The repeated block therefore satisfies
\[
\begin{aligned}
(\sigma_{j+k\ell},\dots,\sigma_{j+(k+1)\ell-1}) &= p_{j,\ell},\\
&\text{for } k=1,\dots,r_{j,\ell}.
\end{aligned}
\]
Only candidates with at least two contiguous repeats are retained. For each retained candidate, the repeated-steps coverage is defined as
\[
c_{j,\ell} = \frac{\ell \, r_{j,\ell}}{n},
\]
and its priority score is
\[
s_{j,\ell} = c_{j,\ell}\left(1+\frac{1}{\ell}\right),
\]
which favors both larger coverage and shorter repeated blocks. Among all retained candidates, the verifier selects the one with the highest score as the dominant loop pattern. A trajectory is then judged inconsistent when this dominant loop covers more than $40\%$ of the trajectory, or when its repeat count is at least four even if the coverage is lower. This criterion captures both short repetitive failures, such as repeated clicking or scrolling, and longer multi-step cycles, while remaining restricted to contiguous repetition rather than scattered recurrence.

\section{Dataset Expansion Details}
\label{sec:data-expand}

After the initial synthesis round, we continue expansion by revisiting transition triplets $(s, op, s')$ extracted from the website map $\mathcal{G}$. The user specifies the number of additional tasks to synthesize, after which the system samples the same number of triplets for a second proposal pass. For each sampled triplet, previously accepted tasks associated with it are prepended to the prompt so that $M_{ui}$ is encouraged to propose a new task that remains compatible with the observed transition while avoiding near-duplicate outputs. Each newly proposed task is then processed by the same collaborative synthesis and post-verification pipeline described in the main text.

The resampling order follows the structural accessibility of $\mathcal{G}$. Let $\operatorname{depth}(s)$ denote the depth of the source state $s$ from the root, and let $\deg(s)$ denote its degree in the website map. We rank triplets lexicographically by
\[
\bigl(\operatorname{depth}(s),\; -\deg(s)\bigr),
\]
so that shallower triplets are revisited first, and among triplets at the same depth, those attached to higher-degree states are preferred. This bias is useful because shallow, high-degree regions usually offer larger interaction space and more stable execution prefixes, making them more likely to yield diverse yet completable tasks.

\section{Experimental Details}
\label{sec:exp}

\subsection{Environment}

\paragraph{Benchmarks and Filtering.} We evaluate on WebArena \citep{zhou2024webarena}, which contains 812 tasks derived from 241 templates across five websites, and WebVoyager \citep{he2024webvoyager}, which contains 643 task queries across 15 websites. For WebArena, we sample one task from each template and remove cross-website tasks following prior work \citep{wang2025adapting}, leaving 226 evaluation tasks. For WebVoyager, we exclude tasks from Google Flights and Booking because they are no longer executable, remove Cambridge Dictionary, Coursera, and Google Search because of strict robot verification, and exclude Google Maps to avoid overlap with the websites used for training in WebArena. This leaves 388 evaluation tasks.

\paragraph{Environment Configuration.} We host the online web environment on an Amazon EC2 instance (\texttt{t3a.xlarge}) with a 1000GB EBS root volume. Both benchmarks provide screenshots and accessibility trees as observations.

\begin{table}[!t]
  \centering
  \footnotesize
  \setlength{\tabcolsep}{3.5pt}
  \begin{tabular}{l p{0.5\linewidth}}
    \toprule[1pt]
    Action & Description \\
    \midrule[0.6pt]
    \texttt{click [id]} & Clicks the element tagged with the given id. \\
    \texttt{type [id] [content]} & Types the provided content into the element tagged with the given id. \\
    \texttt{hover [id]} & Hovers over the element tagged with the given id. \\
    \texttt{press [key comb]} & Presses a key combination (e.g., \texttt{Cmd+V}). \\
    \texttt{scroll [up|down]} & Scrolls the page up or down. \\
    \texttt{new\_tab} & Opens a new browser tab. \\
    \texttt{tab\_focus [tab\_index]} & Switches focus to the browser tab with the specified index. \\
    \texttt{close\_tab} & Closes the current active tab. \\
    \texttt{goto [url]} & Navigates to the specified URL. \\
    \texttt{go\_back} & Navigates to the previously viewed page. \\
    \texttt{go\_forward} & Navigates to the next page. \\
    \texttt{stop [reason]} & Stops execution and records the reason. \\
    \texttt{none [summary]} & Terminates execution with the final answer. \\
    \bottomrule[1pt]
  \end{tabular}
  \caption{Available browsing actions.}
  \label{tab:actions}
\end{table}

\paragraph{Action Space.} Table~\ref{tab:actions} lists all actions available in the trajectories. The \texttt{stop} action terminates execution when a task is infeasible or the agent is stalled. The \texttt{none} action denotes the final execution step and returns the task answer.

\subsection{Baselines}

We compare against NNetNav \citep{murty2024nnetnav}, OS-Genesis \citep{sun2025genesis}, and SynthAgent \citep{wang2025adapting}, and construct matched training sets from their released data.

OS-Genesis provides 1,000 trajectories, which we use in full. For NNetNav and SynthAgent, we sample 1,000 trajectories from each method. To balance the data across websites, we sample 200 trajectories per website. For NNetNav, we first merge multi-step records into single trajectory records according to data indices, and then use regular-expression matching on the first-step text of each trajectory to identify the corresponding website. For SynthAgent, we perform indexed sampling over the task list, whose keys follow the format \texttt{task@website}. Specifically, we select the first 200 task indices for each website and then extract the corresponding trajectories to construct the training dataset.

Because NNetNav and SynthAgent release more trajectories than the 1,000 examples used in our balanced main comparison, we further conduct a supplemental full-data experiment for these two baselines. This experiment fine-tunes the UI-aware model (Qwen3-VL-8B-Instruct) on the full released NNetNav or SynthAgent data using the same LoRA configuration as the main experiments, and evaluates the resulting models on the same WebArena benchmark. Table~\ref{tab:full_data_baselines_webarena} reports the resulting website-level and overall success rates. Using the full released data, NNetNav and SynthAgent achieve overall success rates of 17.26\% and 18.58\%, respectively, improving upon their balanced 1,000-trajectory results of 14.16\% and 16.81\%. Nevertheless, both remain below SynWeaver's 19.91\%, indicating that data quality and structural coverage remain important beyond scale alone.

\begin{table}[t]
  \centering
  \footnotesize
  \setlength{\tabcolsep}{4pt}
  \resizebox{\linewidth}{!}{
  \begin{tabular}{lcccccc}
    \toprule[1pt]
    Method & Shopping & CMS & Reddit & GitLab & Maps & Overall \\
    \midrule[0.6pt]
    NNetNav (full) & 20.00 & 15.79 & 15.38 & 7.14 & 34.38 & 17.26 \\
    SynthAgent (full) & 23.64 & 14.04 & 19.23 & 7.14 & 37.50 & 18.58 \\
    \bottomrule[1pt]
  \end{tabular}}
  \caption{Supplemental full-data baseline results for UI-aware model (Qwen3-VL-8B-Instruct) on WebArena.}
  \label{tab:full_data_baselines_webarena}
\end{table}

\subsection{SynWeaver Data Synthesis}

\begin{table}[t]
  \centering
  \footnotesize
  \setlength{\tabcolsep}{4pt}
  \begin{tabular}{lccccc}
    \toprule[1pt]
    Statistics & Shopping & CMS & Reddit & GitLab & Maps \\
    \midrule[0.6pt]
    States & 90 & 646 & 131 & 426 & 16 \\
    Transitions & 148 & 750 & 203 & 606 & 16 \\
    \bottomrule[1pt]
  \end{tabular}
  \caption{Original website map statistics for WebArena.}
  \label{tab:webarena_map_stats_ori}
\end{table}

\paragraph{Website Map Construction and Filtering.} For the five WebArena websites, we construct a website map using Qwen3-VL-235B-A22B-Instruct to sample representative elements within list containers and Qwen3-VL-Plus as a fallback for state equivalence judgment. Table~\ref{tab:webarena_map_stats_ori} reports the original graph statistics. CMS and GitLab exhibit substantially higher graph complexity, so we apply controlled sampling to balance the final UI and trajectory data. For page-level UI synthesis on these two websites, we retain one state per URL. For transition-level synthesis, we keep one transition every two hops and cap the number of scrolling operations at 25, discarding additional scrolling transitions after the cap is reached. The filtered statistics used for data synthesis are reported in Table~\ref{tab:webarena_map_stats} in the main paper.

\paragraph{Synthesized Data.} Using the retained states, we synthesize two page-level datasets, page description and page QA, with 559 instances each. Using the retained transitions, we synthesize three transition-level datasets, element description, forward transition description, and inverse transition description, with 794 instances each. After shuffling these five subsets, we obtain a UI dataset containing 3,500 entries. We then synthesize one task-trajectory pair for each retained transition. To mitigate sample imbalance, we perform task sampling three times for the Maps website; additional details are provided in Appendix~\ref{sec:data-expand}. This process yields 826 candidate task-trajectory pairs. During post-verification, four pairs fail validation and are removed, leaving 822 validated task-trajectory pairs for training.

\subsection{Training}

We train on a cluster of eight Ascend 910C NPUs. We use Gemini-3-Flash \footnote{https://storage.googleapis.com/deepmind-media/Model-Cards/Gemini-3-Flash-Model-Card.pdf} as the base teacher $M_1$ and Gemini-3.1-Pro \footnote{https://deepmind.google/models/model-cards/gemini-3-1-pro/} as the stronger teacher $M_2$. In the first stage, $M_1$ synthesizes page-level UI data for each retained state and transition-level UI data for each retained transition. We then LoRA fine-tune each backbone to obtain $M_{ui}$ using a learning rate of 5e-5, batch size 16, rank 8, alpha 16, and 3 epochs. In the second stage, we LoRA fine-tune $M_{ui}$ into the final experimental model using the synthesized data of each method, with a learning rate of 1e-4, batch size 32, rank 16, alpha 32, and 3 epochs.

\section{Data Scaling}
\label{sec:data-scaling}

\paragraph{Experimental Setup.}
We investigate how SynWeaver scales with the amount of synthesized task-trajectory supervision. Because each sampled transition in the website map serves as an anchor for task-trajectory synthesis, the number of sampled or resampled transitions directly determines the trajectory budget. We construct four operational data scales, summarized in Table~\ref{tab:data-scaling-config}. The small setting (S) uses half of the standard sample, whereas the medium setting (M) uses the standard sampling configuration. To test whether broader graph coverage provides further gains, the large setting (L) starts from M and exhaustively samples the transitions of Shopping Admin (CMS) and GitLab. Finally, the extra-large setting (XL) augments L with one additional sampling pass whose budget equals that of M. This last setting increases the number of trajectories generated from the available website-map structure and tests whether additional task diversity remains beneficial after expanding transition coverage. Relative to M, the four settings contain $0.50\times$, $1.00\times$, $2.12\times$, and $3.12\times$ as many trajectories, respectively. These points characterize practical scaling behavior under the available website maps rather than fitting a parametric power law.

\begin{table}[H]
  \centering
  \footnotesize
  \setlength{\tabcolsep}{3.5pt}
  \begin{tabular}{c p{0.43\linewidth} r c}
    \toprule[1pt]
    Scale & Data construction & \# Traj. & Rel. \\
    \midrule[0.6pt]
    S  & Half of the standard sample & 413   & $0.50\times$ \\
    M  & Standard sampling configuration & 826   & $1.00\times$ \\
    L  & M with full CMS and GitLab sampling & 1,755 & $2.12\times$ \\
    XL & L plus one additional M-sized sampling pass & 2,581 & $3.12\times$ \\
    \bottomrule[1pt]
  \end{tabular}
  \caption{Construction of the four trajectory scales.}
  \label{tab:data-scaling-config}
\end{table}

For all four settings, we start from the same UI-aware Qwen3-VL-8B-Instruct initialization and keep the synthesis, collaborative refinement, post-verification, fine-tuning hyperparameters, and WebArena evaluation split unchanged. Thus, the comparison varies only the amount and coverage of the task-trajectory supervision induced by the sampling configurations above. We report task success rate (SR) on each website and over all 226 evaluation tasks.

\begin{figure}[H]
  \centering
  \includegraphics[width=\columnwidth]{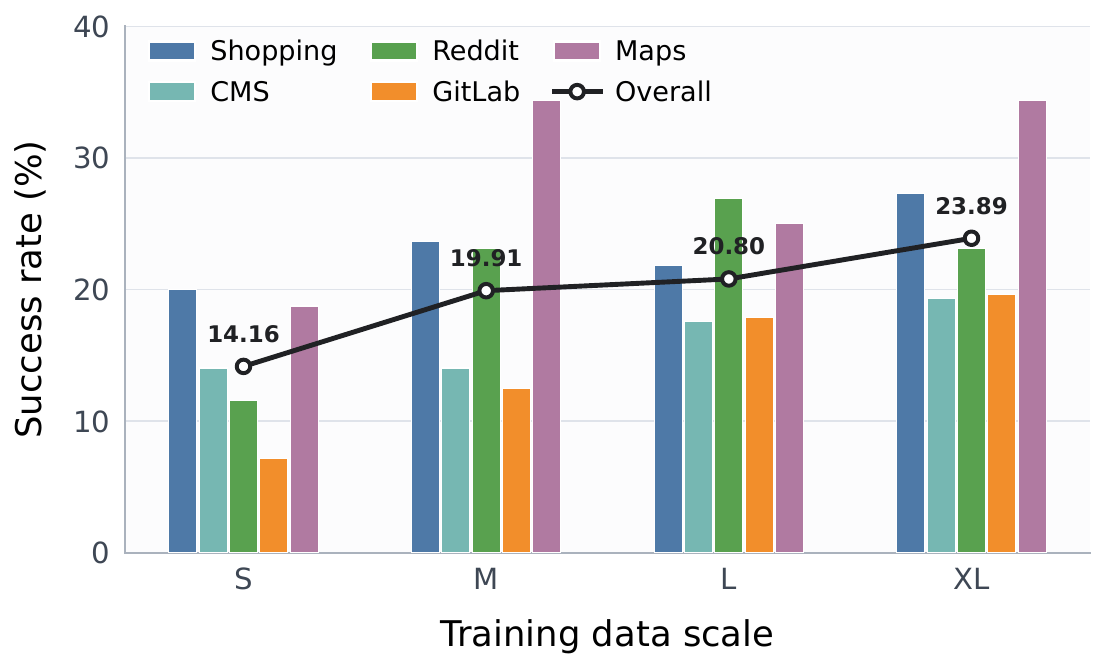}
  \caption{Performance of SynWeaver across different websites with varying data amounts.}
  \label{fig:data-scaling}
\end{figure}

\paragraph{Results.}
Figure~\ref{fig:data-scaling} shows a consistently positive aggregate scaling trend. Increasing the training set from 413 trajectories (S) to 826 trajectories (M) raises the overall SR from 14.16 to 19.91, a gain of 5.75 percentage points. Expanding transition coverage in L further improves the overall SR to 20.80, and the additional sampling pass in XL yields the best result of 23.89. Overall, scaling from S to XL improves SR by 9.73 points, corresponding to a relative improvement of 68.7\%.

The website-level results show a broadly positive but heterogeneous scaling pattern. Although performance on individual websites fluctuates across the intermediate scales, most websites benefit when the data are expanded from S to XL, and the largest setting achieves the strongest or tied-strongest performance on four of the five websites. The non-monotonic changes at intermediate scales likely reflect shifts in data composition: L primarily expands transition coverage for CMS and GitLab, whereas XL further increases task diversity through an additional sampling pass. Taken together with the steadily increasing overall SR, these results indicate that SynWeaver can effectively exploit additional website-map coverage and repeated transition-conditioned sampling, while the distribution of the synthesis budget across websites remains important for domain-specific gains.

\section{UI Test}
\label{sec:ui-test}

\paragraph{Evaluation Data Construction.}
To determine whether website-prior learning improves the model's understanding of website-specific UI knowledge, we construct a held-out UI question-answering benchmark from online interactions in WebArena. We first run the teacher model, Gemini-3-Flash, on all 226 WebArena evaluation tasks and record each interaction step in real time. For every step, we retain the screenshots before and after the interaction, the executed action, and the target element on which the action is performed. We then randomly sample 200 unique interaction steps, subject to the constraint that none of them appears in the website maps used to construct the UI training data. This separation prevents direct overlap between the UI training supervision and the evaluation examples.

Following the evaluation protocol of GUI Knowledge Bench \citep{shi2025gui}, we use the sampled interactions to construct four categories of UI knowledge questions: \textbf{Widget Function}, which evaluates whether the model understands the purpose of a UI element; \textbf{Layout Semantics}, which evaluates its understanding of the element's role and spatial relationship within the page layout; \textbf{Interaction Effect}, which asks the model to predict or identify the state change caused by an interaction; and \textbf{Interaction Type}, which tests whether the model can infer the appropriate operation for a state change. Each category contains 50 questions generated from a disjoint subset of 50 interactions. Consequently, the benchmark contains 200 questions in total, with each sampled interaction contributing to exactly one question.

\paragraph{Models and Metric.}
We evaluate the base Qwen3-VL-8B-Instruct model, its UI-aware counterpart obtained after website-prior learning (+UI), and Gemini-3.1-Pro. We report question-answering accuracy over all 200 examples.

\begin{table}[H]
  \centering
  \small
  \begin{tabular}{l c}
    \toprule[1pt]
    Model & Accuracy (\%) \\
    \midrule[0.6pt]
    Qwen3-VL-8B-Instruct & 74.5 \\
    \quad+UI & \textbf{89.0} \\
    Gemini-3.1-Pro & 87.5 \\
    \bottomrule[1pt]
  \end{tabular}
  \caption{Accuracy on the held-out UI knowledge test.}
  \label{tab:ui-test}
\end{table}

\paragraph{Results.}
As shown in Table~\ref{tab:ui-test}, website-prior learning improves Qwen3-VL-8B-Instruct from 74.5\% to 89.0\%, an absolute gain of 14.5 percentage points. The resulting UI-aware model also exceeds Gemini-3.1-Pro by 1.5 points on this test. Because all evaluation interactions are unique and excluded from the website maps used for UI data construction, the improvement cannot be attributed to direct memorization of the tested transitions. Instead, the results indicate that the UI supervision enables the model to acquire transferable knowledge on the target websites.

\section{UI Dataset Cases}

This appendix shows five types of UI data: page description, page QA, element description, forward transition description, and inverse transition description.

\label{sec:ui-data}

\begin{figure*}[t]
  \centering
  \includegraphics[width=\textwidth]{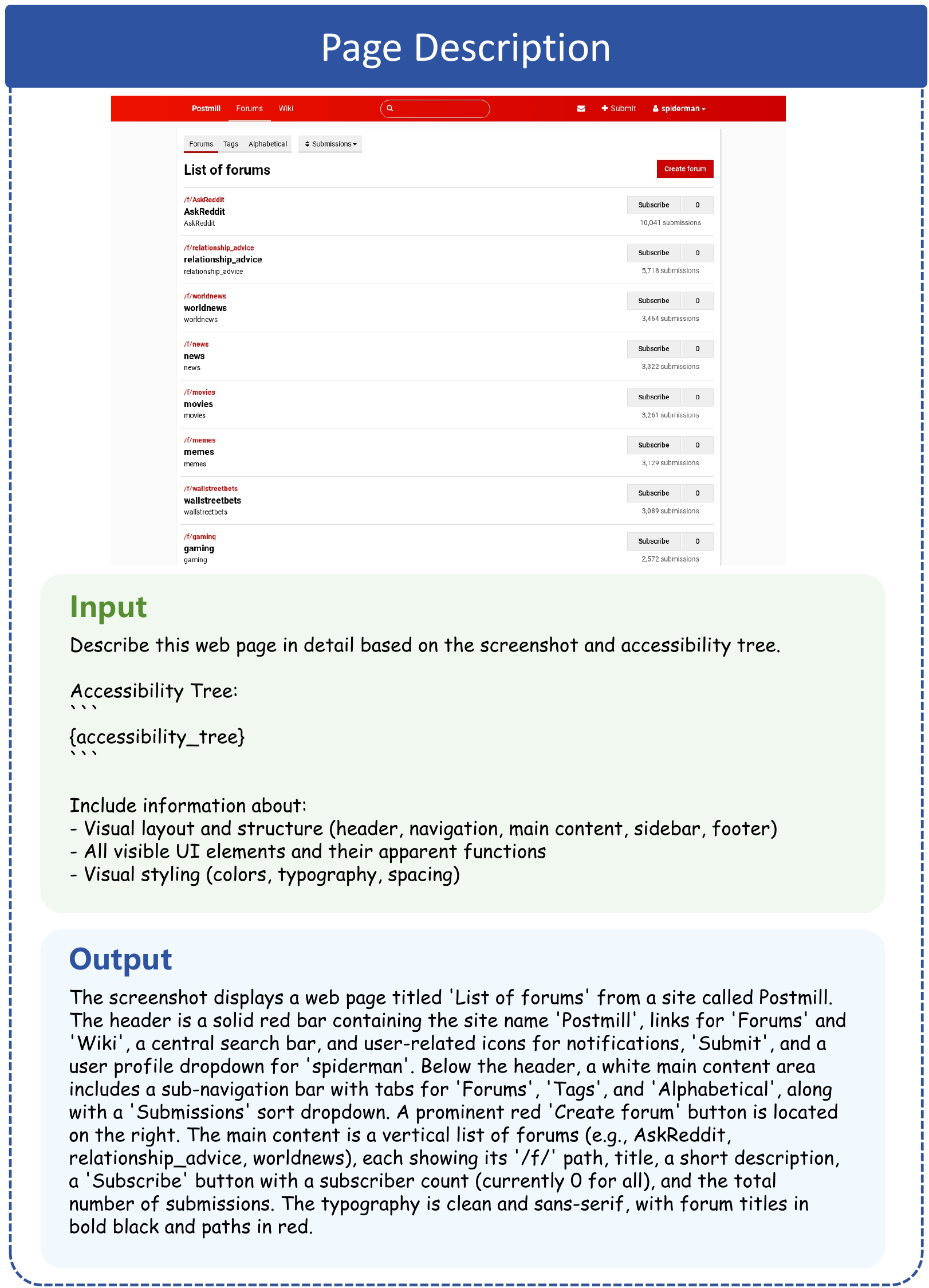}
  \caption{Example of page description data in the UI dataset.}
  \label{fig:ui-page-desc}
\end{figure*}

\begin{figure*}[t]
  \centering
  \includegraphics[width=\textwidth]{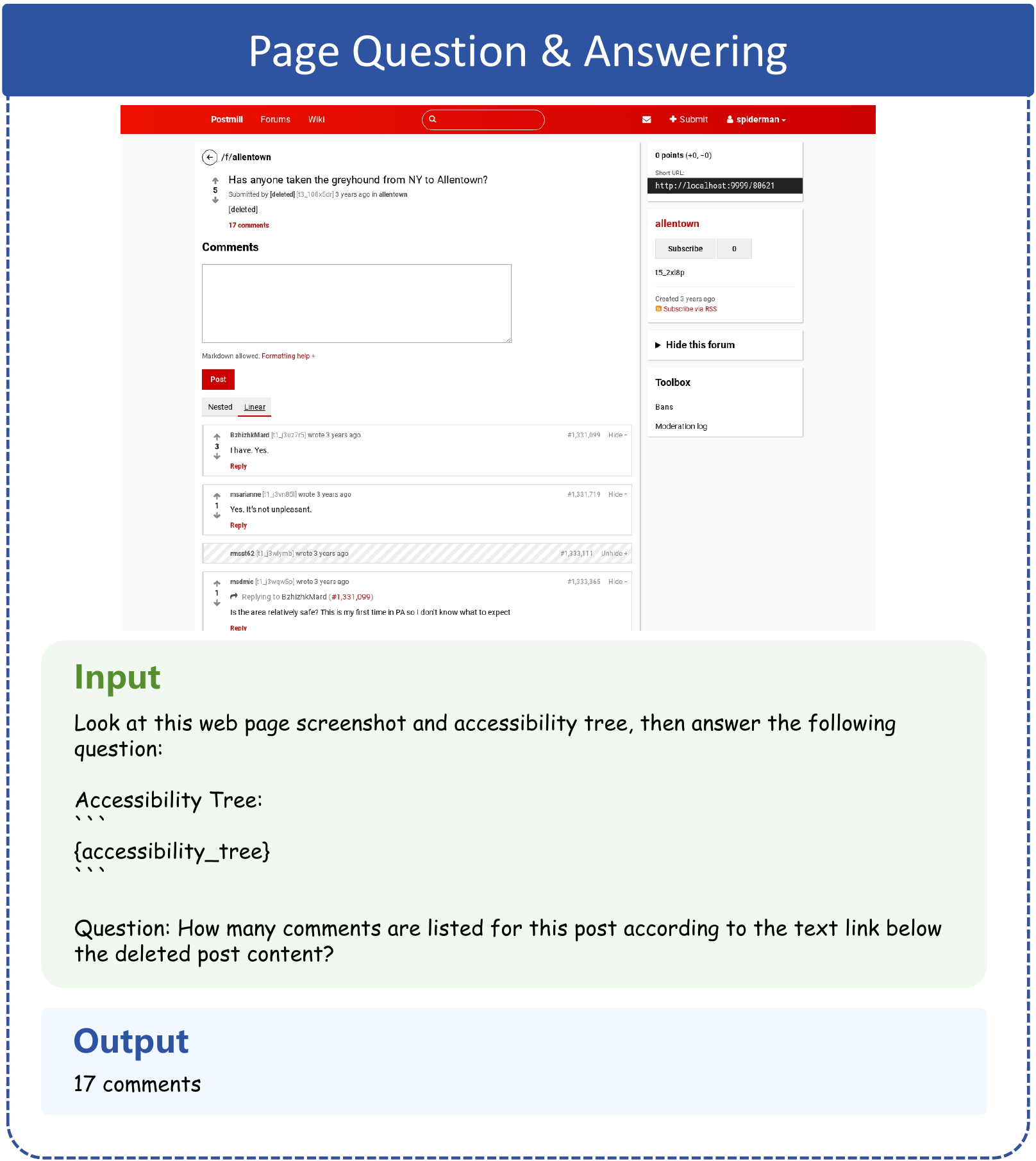}
  \caption{Example of page QA data in the UI dataset.}
  \label{fig:ui-page-qa}
\end{figure*}

\begin{figure*}[t]
  \centering
  \includegraphics[width=\textwidth]{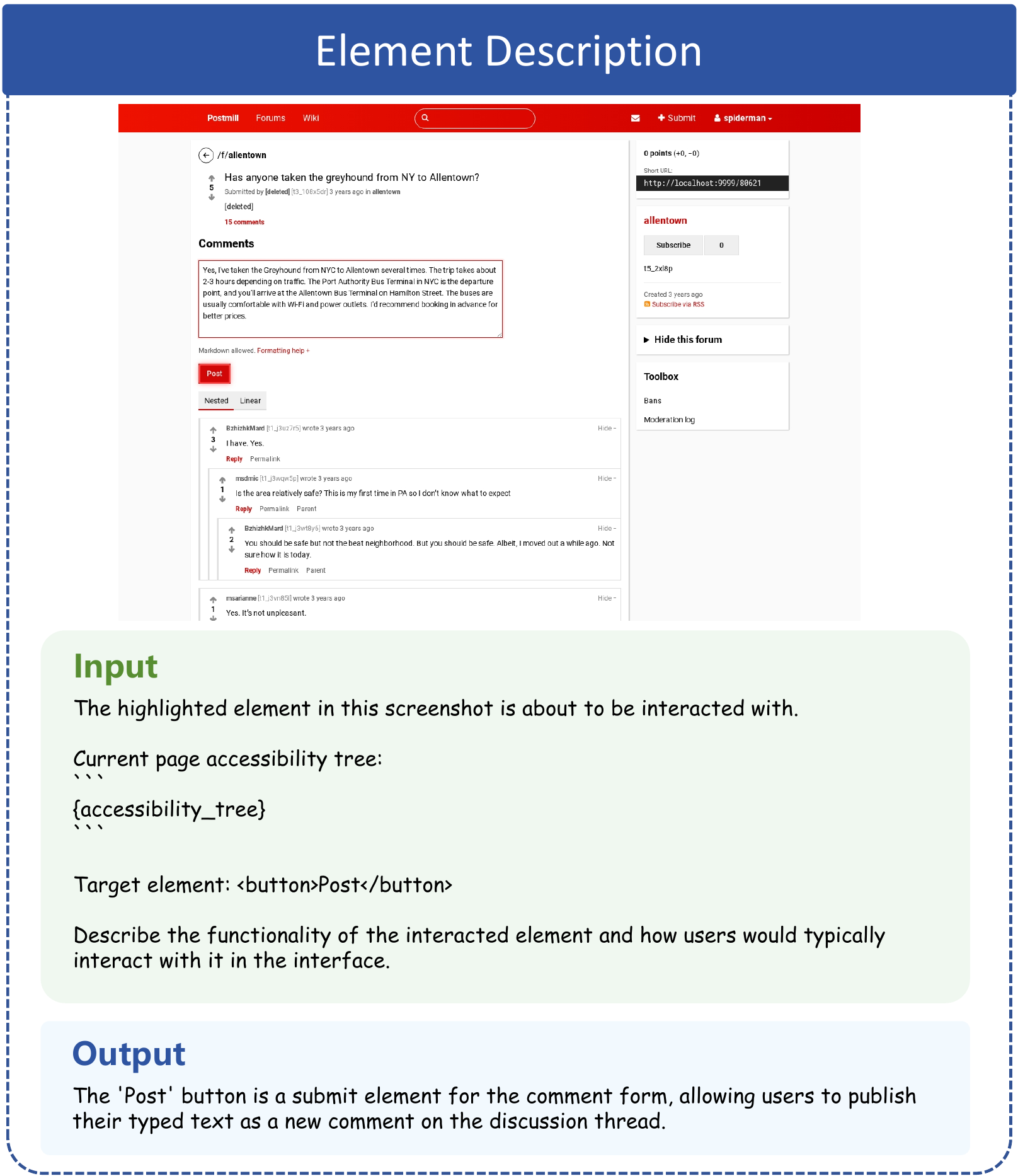}
  \caption{Example of element description data in the UI dataset.}
  \label{fig:ui-elem-desc}
\end{figure*}

\begin{figure*}[t]
  \centering
  \includegraphics[width=\textwidth]{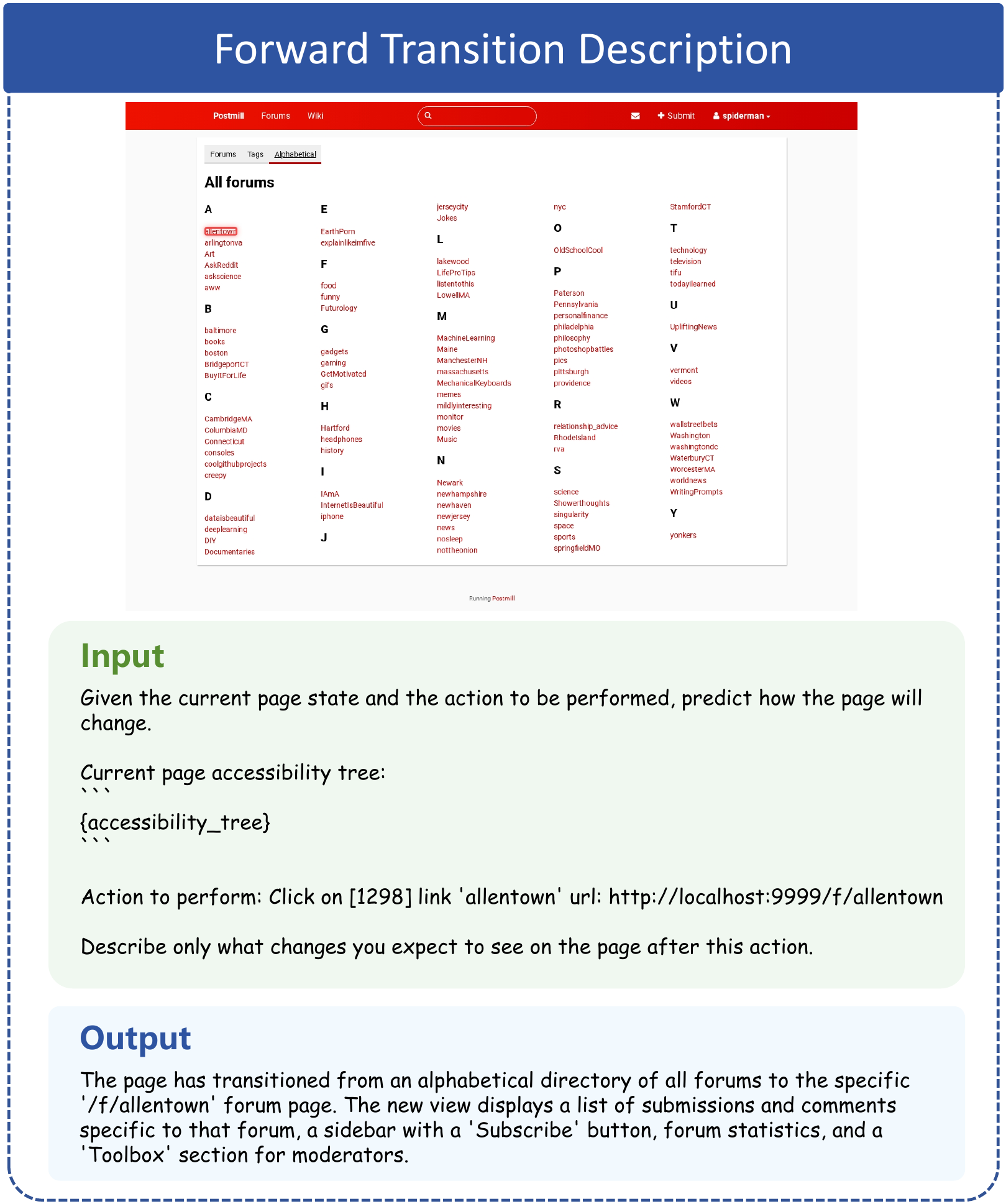}
  \caption{Example of forward transition description data in the UI dataset.}
  \label{fig:ui-forward-desc}
\end{figure*}

\begin{figure*}[t]
  \centering
  \includegraphics[width=\textwidth]{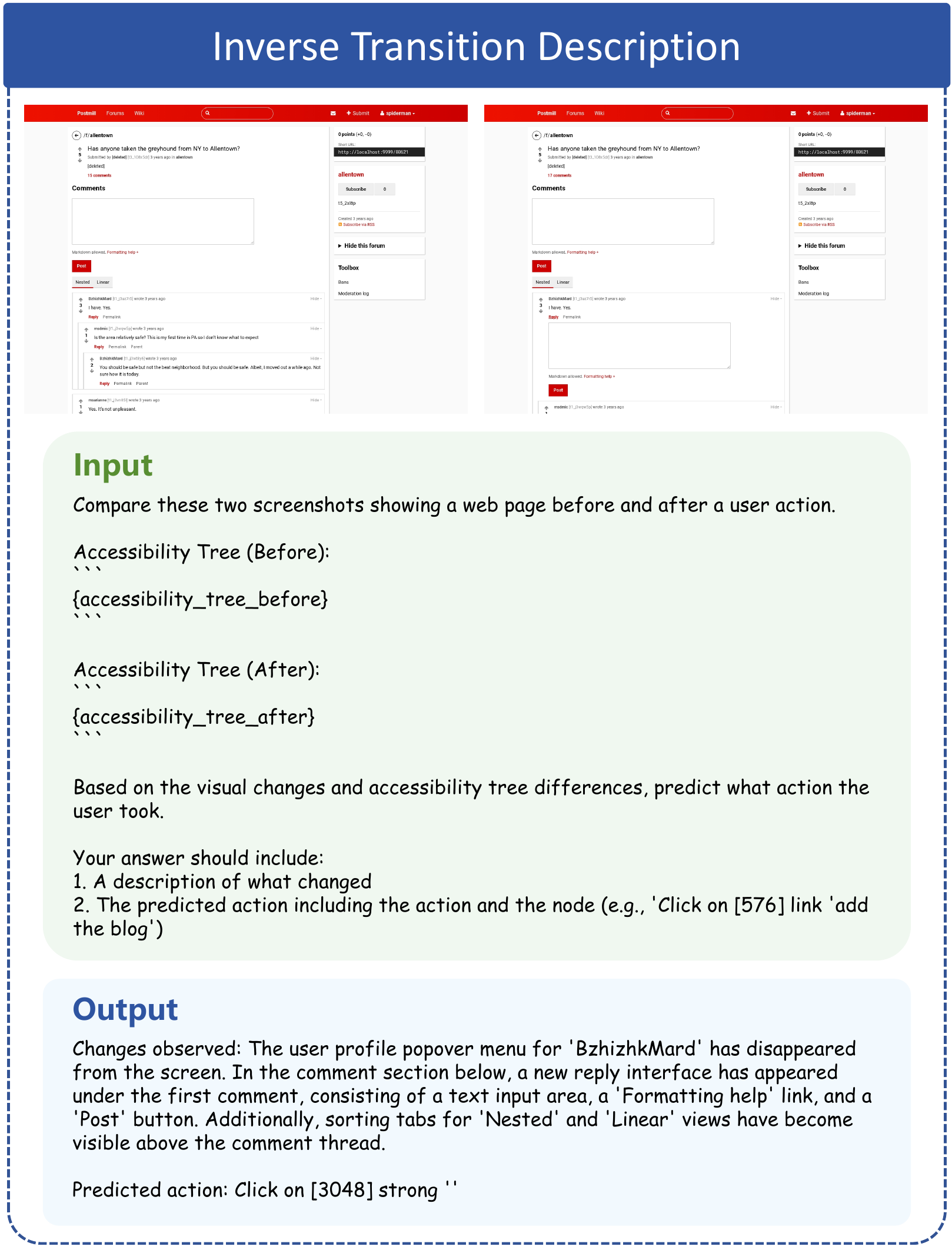}
  \caption{Example of inverse transition description data in the UI dataset.}
  \label{fig:ui-inverse-desc}
\end{figure*}

\clearpage
\section{Prompts}
\label{sec:prompts}

This appendix provides the main prompts used in SynWeaver.

\newtcblisting{promptbox}{
    breakable,
    width=\linewidth,
    colback=gray!8,
    colframe=black,
    boxrule=0.8pt,
    arc=0pt,
    left=2pt,
    right=2pt,
    top=2pt,
    bottom=2pt,
    before skip=0.75\baselineskip,
    after skip=0.5\baselineskip,
    listing only,
    listing options={
        basicstyle=\ttfamily\scriptsize,
        breaklines=true,
        breakautoindent=false,
        breakindent=0pt,
        breakatwhitespace=false,
        columns=fullflexible,
        keepspaces=true,
        showstringspaces=false,
        aboveskip=0pt,
        belowskip=0pt,
    },
}

\begin{promptbox}
You are a webpage structure analysis model.  
Your task is to determine whether the elements inside an HTML list (such as `<ul>` or `<ol>`) represent **repeated functionality** (only one element or one group of elements needs to be clicked) or **distinct functionality** (all elements must be clicked).

You will be given the textual structure of a list, including each element's index and text content.  
Your goal is to analyze the semantics and structural patterns to decide whether the list items lead to **similar** or **different** functional outcomes.

## Objective

1. **If the list items represent repeated or symmetric functionality**: Only **one** representative item or **one group of items** should be clicked.

2. **If the list items represent distinct functionality**: All items must be clicked.

## Classification Criteria

### **1. Treat the items as *functionally identical* (decision = "single") if:**

All or most items exhibit these characteristics:

- Text content represents members of the same category (e.g., product categories, product titles, tags, results)
- Elements have a highly repetitive structure (e.g., repeated `<li>`, repeated `<a>`, repeated `<div>` templates)
- Items correspond to the same type of resource (e.g., products, categories, search results)
- Clicking each item typically leads to pages of the **same structure but different content** (e.g., different product detail pages)
- The list visually or semantically presents a "set of options of the same type"

If these conditions are met, treat the list as **repeated functionality**.

### **2. Treat the items as *functionally different* (decision = "all") if:**

Any of the following are true:

- Text content clearly indicates different action targets (e.g., "new communication", "Sign In", "search communication", "profile")
- Items represent different task categories (e.g., login, registration, account management)
- The list acts as a navigation menu with distinct entry points  
- Clicking each item evidently results in **different functionality**, not symmetrical options

If any such condition applies, treat the list as **distinct functionality**.

## Output Format

You must output a JSON object in the following format:
```json
{
  "decision": "single" or "all",
  "reason": "Explanation of the decision",
  "selected_indices": when the decision is "single", return [list of indices]. Otherwise, return an empty list.
}
```

## Example

### Example 1

Input:
```
<ul>
[21]<a> Black Friday </a>
[22]<a> Amazon Haul </a>
[23]<span> Medical Care </span>
[24]<button>  </button>
[25]<a> Best Sellers </a>
[26]<a> Amazon Basics </a>
[27]<a> New Releases </a>
[28]<a> Books </a>
[29]<a> Registry </a>
[30]<span> Groceries </span>
[31]<button>  </button>
[32]<span> Gift Cards </span>
</ul>
```

Output:
```json
{
  "decision": "all",
  "reason": "The list items represent distinct functional entry points rather than repeated or symmetric options. Items such as 'Black Friday', 'Amazon Haul', 'Medical Care', 'Best Sellers' all lead to different content categories or features. They do not share a repeated structure or identical functional behavior. Therefore, each item must be treated as a unique function.",
  "selected_indices": []
}
```

### Example 2

Input:
```
<ul>
[12]<img>  </img>
[13]<span> Outstanding Fiction of the past 25 years </span>
[14]<img>  </img>
[15]<span> Unforgettable Nonfiction of the past 25 years </span>
[16]<img>  </img>
[17]<span> Most buzzworthy books of the past 25 years </span>
[18]<img>  </img>
[19]<span> Editors' personal favorites </span>
</ul>
```

Output:
```
{
  "decision": "single",
  "reason": "The list items represent a set of thematically similar options: each entry pairs an image with a descriptive label for a specific curated book collection. The structure is repetitive, and clicking any item is expected to lead to a page of the same functional type (a book collection page), differing only in content. Therefore, these items exhibit repeated functionality.",
  "selected_indices": [12, 13]
}
```
\end{promptbox}

\begin{center}
\begin{minipage}{\linewidth}
\captionsetup{hypcap=false}
\captionof{prompt}{Prompt for sampling representative items from the list container.}
\label{prompt:sample-item}
\end{minipage}
\end{center}

\begin{promptbox}
You are an intelligent form filling assistant. Your task is to generate reasonable test values for each input field in a web form based on the form context and field semantics.

## Context

You will receive:
1. **Page URL**: The current page URL to help understand the context
2. **Accessibility Tree**: A structured representation of the page content around the form
3. **Form Fields**: A list of input fields in the form, each with:
   - Index number (1, 2, 3, ...)
   - Tag type (input/textarea/select)
   - Input type (for input elements: text, email, password, etc.)
   - Name attribute
   - Aria-label
   - Placeholder text
   - Current text content

## Your Task

Generate appropriate test values for each form field. The values should be:
1. **Semantically appropriate**: Match the expected content type (email for email fields, phone for phone fields, etc.)
2. **Realistic**: Look like real user input, not obviously fake
3. **Valid**: Pass basic validation (proper email format, valid phone format, etc.)
4. **Contextual**: Consider the form's purpose based on URL and accessibility tree

## Output Format

Return a JSON object with a "values" array containing the fill value for each field **in order**:

```json
{
  "values": ["value_for_field_1", "value_for_field_2", "value_for_field_3", ...],
  "reasoning": "Brief explanation of the values chosen"
}
```

**Important**:
- The array length MUST match the number of input fields
- Each value corresponds to the field at that index position
- For select elements, provide the option text to select
- For password fields, use a strong test password like "TestPass123!"
- For date fields, use ISO format (YYYY-MM-DD)

## Examples

### Example 1: Login Form

Input:
- URL: https://example.com/login
- Fields:
  1. input[type=email] name="email" placeholder="Enter your email"
  2. input[type=password] name="password" placeholder="Password"

Output:
```json
{
  "values": ["testuser@example.com", "TestPass123!"],
  "reasoning": "Standard login form with email and password fields"
}
```

### Example 2: Registration Form

Input:
- URL: https://shop.example.com/register
- Fields:
  1. input[type=text] name="first_name" aria-label="First Name"
  2. input[type=text] name="last_name" aria-label="Last Name"
  3. input[type=email] name="email"
  4. input[type=tel] name="phone" placeholder="Phone number"

Output:
```json
{
  "values": ["John", "Smith", "john.smith@example.com", "555-123-4567"],
  "reasoning": "E-commerce registration form requiring personal contact information"
}
```
\end{promptbox}

\begin{center}
\begin{minipage}{\linewidth}
\captionsetup{hypcap=false}
\captionof{prompt}{Prompt for filling the form container.}
\label{prompt:fill-form}
\end{minipage}
\end{center}

\begin{promptbox}
You are an expert web page analyzer. Your task is to determine whether two screenshots represent the **same functional state** of a web page, even if some minor visual elements differ.

# Background

During web crawling and state replay, certain transient UI elements (like toast notifications, temporary alerts, or dynamic content) may appear or disappear. However, the underlying page state remains the same. We need to identify whether two screenshots represent the same functional page state while ignoring these superficial differences.

# Task

Compare the two provided screenshots and determine if they represent the **same web page state**.

# What to Consider as the SAME State

Two screenshots should be considered the **same state** if they share:

1. **Core Page Structure**:
   - Same page layout and main content areas
   - Same navigation elements (header, footer, sidebar)
   - Same primary content sections

2. **Functional Elements**:
   - Same interactive elements (buttons, links, forms, input fields)
   - Same product listings, article content, or data tables
   - Same menu items and navigation options

3. **Page Identity**:
   - Same URL path (if visible)
   - Same page title or heading
   - Same main purpose/function

# What to IGNORE (Acceptable Differences)

The following differences should be **ignored** when comparing states:

1. **Transient Notifications**:
   - Toast messages (success/error/warning notifications)
   - Temporary alert boxes
   - Popup hints or tooltips
   - Tutorial/onboarding overlays
   - Cookie consent banners

2. **Dynamic Content**:
   - Timestamps (e.g., "2 minutes ago" vs "5 minutes ago")
   - Real-time counters (shopping cart quantity, notification badges)
   - Live data feeds (stock prices, weather updates)
   - Advertisement content
   - Personalized recommendations

3. **Visual State Changes**:
   - Hover effects or focus states
   - Loading spinners or progress indicators
   - Scroll position
   - Animation states
   - Dropdown menu open/closed states

4. **Session-Specific Elements**:
   - Session IDs or tokens in URLs
   - CSRF tokens
   - Temporary promotional banners
   - A/B testing variations (minor UI tweaks)

# Decision Criteria

- **SAME**: If the core page structure, main content, and functional elements are identical (ignoring the acceptable differences listed above)
- **DIFFERENT**: If there are substantial differences in layout, content, navigation, or functionality

# Output Format

Provide your analysis in the following JSON format:

```json
{
  "decision": "same" or "different",
  "confidence": 0.0 to 1.0,
  "reasoning": "Simple explanation of your decision",
}
```
\end{promptbox}

\begin{center}
\begin{minipage}{\linewidth}
\captionsetup{hypcap=false}
\captionof{prompt}{Prompt for state comparison.}
\label{prompt:compare-state}
\end{minipage}
\end{center}

\begin{promptbox}
You are an expert UI analyst tasked with analyzing web page screenshots and accessibility trees. Your analysis will be used to train AI models to understand web interfaces.

You will receive:
1. A screenshot of a web page
2. The page's accessibility tree

Your task is to provide TWO types of analysis:

## Part 1: Page Description
Provide a comprehensive description of the page that includes:
- Visual layout and structure (header, navigation, main content, sidebar, footer)
- All visible UI elements and their apparent functions
- Visual styling (colors, typography, spacing)

## Part 2: Page QA
Generate ONE informative question-answer pair about the page. The question should:
- Ask about specific information visible on the page
- Be answerable ONLY by looking at the screenshot
- Focus on extracting factual information (not opinions)
- Examples: "What is the main heading?", "How many products are displayed?", "What navigation options are available?"

## Output Format
You MUST respond in the following JSON format:
```json
{
    "page_description": "Your detailed page description here...",
    "question": "Your question about the page content...",
    "answer": "The answer based on the screenshot..."
}
\end{promptbox}

\begin{promptbox}
Analyze the following web page:

URL: {url}

Accessibility Tree:
```
{accessibility_tree}
```

Please provide:
1. A detailed description of this page
2. One question-answer pair about specific information visible on the page

Remember to respond in the specified JSON format.
\end{promptbox}

\begin{center}
\begin{minipage}{\linewidth}
\captionsetup{hypcap=false}
\captionof{prompt}{System prompt and user prompt for page-level UI data generation.}
\label{prompt:page-ui-data}
\end{minipage}
\end{center}

\begin{promptbox}
You are an expert UI analyst tasked with analyzing web page interactions. Your analysis will be used to train AI models to understand how UI elements work and how pages change after user actions.

You will receive:
1. A "before" screenshot showing the page with a marked element
2. An "after" screenshot showing the page after the action was performed
3. Information about the action taken

Your task is to provide TWO types of analysis:

## Part 1: Element Description
Describe the functionality of the interacted element and how users would typically interact with it in the interface.

## Part 2: State Transition Description
Describe the changes between the before and after states.

IMPORTANT: In the state transition description, describe ONLY what changed on the page. Do NOT mention the action taken (click, fill, etc.) - focus purely on the observable differences between the two screenshots.

## Output Format
You MUST respond in the following JSON format:
```json
{
    "element_description": "Description of the element's functionality...",
    "state_transition": "Description of how the page changed (without mentioning the action)..."
}
\end{promptbox}

\begin{promptbox}
Analyze this web page interaction:

Action Performed: {action}
Element: {element}
URL Before: {from_url}
URL After: {to_url}

The first image shows the page BEFORE the action (with the target element highlighted).
The second image shows the page AFTER the action was performed.

Please provide:
1. A description of the element's functionality
2. A description of how the page changed (without mentioning what action was taken)

Remember to respond in the specified JSON format.
\end{promptbox}

\begin{center}
\begin{minipage}{\linewidth}
\captionsetup{hypcap=false}
\captionof{prompt}{System prompt and user prompt for transition-level UI data generation.}
\label{prompt:trans-ui-data}
\end{minipage}
\end{center}

\begin{promptbox}
You are a GUI (Graphical User Interface) expert capable of analyzing interface changes and envisioning executable tasks or instructions. Given a GUI interface change caused by an action (e.g., clicking or typing) and the corresponding element highlighted in red boxes, you are required to analyze the interface and generate related tasks.

Your task is to envision tasks based on the current action and the resulting changes in the screenshots. The output should include three components:

1. Sub-Instruction: Create a natural language instruction for the current action based on the interface changes it caused. The instruction should be concise, clear, and actionable, incorporating specific details critical to the task, such as elements, file names, timestamps, or other relevant content visible in the screenshots. For example:
   - "Click on the 'Add to Cart' button next to the product to add it to your shopping cart."
   - "Type 'OpenAI' into the search bar to find relevant articles."
   - "Scroll down to view the latest blog posts on the homepage."

2. Analysis: Carefully analyze the before-and-after screenshots step by step, focusing on the changes caused by the action. Then, examine key elements in both screenshots and consider possible operations based on these elements. For example: "The previous screen displayed the main interface of a shopping website, featuring multiple product categories and several showcased items. After clicking the 'Sign Up' button, the interface transitioned to a login page where an email and password can be entered to log into an account. The login page also provides other options, such as recovering a password, creating a new account, or logging in with a Google account."

3. High-Level Instruction: Based on the before-and-after screenshots, the action, and the analysis, generate a high-level task that you believe can be completed within the current interface. There are three types of tasks:
   - Information seeking: The user wants to obtain certain information from the webpage, such as product details, reviews, map information, or route comparisons. Please propose clear and specific questions that need an explicit answer, and avoid asking for summary-type questions, such as "summarize the information about a product."
   - Site navigation: The user wants to navigate to a specific page or state.
   - Content modification: The user wants to modify the content of a webpage or its settings.

The high-level instruction should be creative. You need to deeply analyze the elements and executable actions on the interface to generate realistic, valuable, and executable tasks that can be completed within the current GUI. The instruction should be specific, actionable, and goal-oriented, ensuring the task can be completed on the current GUI by including all critical specifics such as file names, relevant timings, or required details.

Below is a brief description of the current website: {website_intro}

Here are some examples of High-Level Instruction for reference:
{task_examples}

Current Action: {current_action_str}

Website Name: {website_name}

Before-action Screenshot: <image is provided in the first attachment> (the action's target element is highlighted in red box if applicable)

After-action Screenshot: <image is provided in the second attachment>

Please generate tasks that can be completed on the current platform, and avoid tasks that are unrelated to the current website.

You MUST respond in the following JSON format: (no extra commentary):
```json
{{
  "Sub-Instruction": "xxx",
  "Analysis": "xxx",
  "High-Level-Instruction": "xxx"
}}
```
RETURN ONLY THE JSON I ASKED FOR.
\end{promptbox}

\begin{center}
\begin{minipage}{\linewidth}
\captionsetup{hypcap=false}
\captionof{prompt}{Prompt for task synthesis, adopted from OS-Genesis \citep{sun2025genesis} and SynthAgent \citep{wang2025adapting}.}
\label{prompt:task-synth}
\end{minipage}
\end{center}

\begin{promptbox}
You are a GUI (Graphical User Interface) Web Agent expert capable of long-horizon planning and executing high-level tasks on a website. Based on the observations and the high-level task to complete, generate the next low-level instruction.

**Information**

1. High-Level Task (your ultimate goal to finish):
"{high_level_task}"

2. History of Actions ({hint_for_history}):
{previous_actions}

3. Current Page (only current view, not full page, you may need to scroll to see more):
    - URL:
      {url}

    - Accessibility Tree (Page Context):
      {page_context}
      
    - Screenshot (only current view, not full page):
      {img_info}
      
**Available Low-Level Actions (exact JSON formats)**

Page Operation Actions:
- CLICK:
  {{"type": "CLICK", "element_id": <int>, "value": ""}}

- TYPE (default behavior is to press Enter after typing unless you explicitly set press_enter_after to 0):
  {{"type": "TYPE", "element_id": <int>, "value": "text to type <string>"}}

- HOVER:
  {{"type": "HOVER", "element_id": <int>, "value": ""}}

- PRESS (keyboard shortcut. The value MUST be either exactly ONE key, e.g., ArrowDown, or ONE key chord joined by "+", e.g., Ctrl+V / Cmd+V):
  {{"type": "PRESS", "element_id": "", "value": "key_comb <string>"}}

- SCROLL:
  {{"type": "SCROLL", "element_id": "", "value": "up" or "down"}}

Tab Management Actions:
- NEW_TAB (open a new, empty browser tab):
  {{"type": "NEW_TAB", "element_id": "", "value": ""}}

- TAB_FOCUS (switch focus to a specific tab by its index, starting from 0):
  {{"type": "TAB_FOCUS", "element_id": "", "value": "<tab_index as int>"}}

- CLOSE_TAB (close the currently active tab):
  {{"type": "CLOSE_TAB", "element_id": "", "value": ""}}

URL Navigation Actions:
- GOTO (navigate directly to a URL):
  {{"type": "GOTO", "element_id": "", "value": "url <string>"}}

- GO_BACK:
  {{"type": "GO_BACK", "element_id": "", "value": ""}}

- GO_FORWARD:
  {{"type": "GO_FORWARD", "element_id": "", "value": ""}}

Completion Actions:
- NONE (use ONLY when the task is completed and you have the final answer):
  {{"type": "NONE", "element_id": "", "value": "summary content or final answer <string>"}}

- STOP (use ONLY when the task is impossible to complete):
  {{"type": "STOP", "element_id": "", "value": "the reason for the STOP <string>"}}

**Critical Rules for Success**
1. You must issue only actions that are valid given the current observation (accessibility tree and screenshot).
2. Only propose ONE atomic action; actions must be executable independently.
3. Prefer actions grounded by element IDs present in the accessibility tree when using CLICK/TYPE/HOVER.
4. You MUST provide meaningful and non-empty value if the action type is in {{TYPE, SCROLL, GOTO, NONE, STOP, TAB_FOCUS}}.
5. When you believe the task is complete (e.g., you have the answer), use NONE with the final answer in value.
6. Be concise and avoid redundant or risky actions; ensure each action clearly advances the task.
7. Use STOP only when:
   - The task lacks necessary information
   - The target does not exist (hallucination)
   - The task is harmful or inappropriate
   - Multiple attempts (>=3) have failed to make progress

8. You MUST first generate a "state_observation_summary" to observe the current environment, then take a step-by-step "reasoning" to decide the next action.
9. The high-level task often requires MULTIPLE steps to complete. Do NOT expect to finish in a single action.
10. You MUST actively decide the next step. Do NOT choose "NONE" or "STOP" unless you are sure the task is finished or impossible.
11. Choose element IDs from the accessibility tree and use them directly in CLICK/TYPE/HOVER actions.
12. If the page does not change after an action, try scrolling to see more elements.
13. When typing dates, use the format "MM/DD/YYYY".

**Output Requirements**
You MUST return a JSON dictionary with the following format (no extra commentary):
{{
    "state_observation_summary": "your 1-3 sentence summary of the current state relevant to the task",
    "reasoning": "your step-by-step reasoning to decide the next action",
    "next_action": {{
        "action": {{"type": "XXXX", "element_id": <int or "">, "value": <string or "">}}
    }}
}}
RETURN ONLY THE DICTIONARY WITHOUT ANY COMMENTARY.
\end{promptbox}

\begin{center}
\begin{minipage}{\linewidth}
\captionsetup{hypcap=false}
\captionof{prompt}{Prompt for task execution on the website.}
\label{prompt:task-exec}
\end{minipage}
\end{center}

\begin{promptbox}
You are a GUI Web Agent expert specializing in task and trajectory optimization. The execution agent has encountered a STOP action, indicating it cannot complete the current task with the existing trajectory.

**Your Primary Goal**: Refine the task description so the execution agent can overcome its current obstacle and complete the task within 2-3 additional steps.

## Current Situation

**High-Level Task**:
"{current_task}"

**Previous High-Level Tasks Attempted**:
{previous_tasks}

**Execution Trajectory** (steps 1-{total_steps}):
{previous_actions}

**Current Page URL**:
{current_url}

**Current Page Context (Accessibility Tree)**:
{page_context}

**Screenshot**:
{img_info}

**Important Note**: The last step is a STOP action which produced no actual operation and will be automatically discarded. Focus your optimization on making the trajectory effective before that point.

---

## Optimization Strategies

Your goal is to create a refined task that enables the agent to complete its objective within 2-3 more steps. Choose between:

### Strategy 1: Task-Only Optimization
**Use when**: The executed trajectory represents meaningful progress, but the original task was too ambiguous or unachievable given the current context.

**Goal**: **Clarify user intent and align with the current trajectory, BUT DO NOT describe steps.**
The refined task should focus on *WHAT* to achieve (the goal state), not *HOW* to do it (the actions).

**Example**: 
- Bad (Instructional): "Click the blue 'Add to Cart' button and then proceed to checkout." (Too specific, loses intent)
- Good (Intent-Driven): "Add the selected item to the shopping cart." (Clear goal, lets agent decide actions)

**How to apply**:
- Change the original intent if necessary to fit the trajectory's progress.
- Clarify the *target entity* or *specific condition* required for success.
- Keep `step_order` as the original order [{original_order}] (excluding the STOP step)
- Set `modified_reasonings` to empty dict {{}}

### Strategy 2: Task-Trajectory Co-Optimization  
**Use when**: The trajectory contains redundant, incorrect, or out-of-order steps that prevent task completion, even with intent clarification.

**Example scenario**: Agent clicked wrong tabs, navigated back and forth, or executed steps in illogical order.

**How to apply**:
- Refine the task description to strictly reflect the core user intent.
- Change the original intent if necessary to fit the trajectory's progress and make it achievable in 2-3 steps.
- Remove or reorder trajectory steps to create a coherent path:
  - **Delete**: Omit step numbers (e.g., [1, 3, 5] removes steps 2, 4)
  - **Reorder**: Change sequence (e.g., [1, 4, 3, 5] swaps steps 3 and 4)
  - **Combined**: [1, 5, 6] means delete steps 2-4 and keep rest
- Update reasoning for affected steps in `modified_reasonings` to ensure logical flow

**Critical**: After modifications, the trajectory must be coherent:
- Each step should logically follow the previous one
- The reasoning should reflect the new context (e.g., "Since we removed the navigation step, now proceed directly to...")
- The modified trajectory should lead naturally to task completion in 2-3 more steps

---

## Decision Framework

Ask yourself these questions in order:

1. **Root Cause Analysis**: Why did the agent issue a STOP action? 
   - Task was impossible/invalid?
   - Task was unclear/ambiguous?
   - Trajectory led to wrong page/state?
   - Agent got stuck in a loop?

2. **Can Task-Only Optimization Succeed?**
   - Does the current page state enable task completion?
   - Will clarifying or changing the **goal state** guide the agent to success in 2-3 steps?
   - Are the executed steps (before STOP) generally correct?
   
   - If YES, use Strategy 1 (Task-Only).
   - If NO, proceed to Question 3.

3. **What Trajectory Issues Exist?**
   - Are there redundant/unnecessary steps?
   - Are steps in the wrong order?
   - Did early mistakes prevent success?
   
   - If YES, use Strategy 2 (Co-Optimization).

4. **Coherence Check** (for Strategy 2):
   - After reordering/deleting, does the trajectory tell a logical story?
   - Do you need to update reasoning to maintain coherence?
   - Will the modified trajectory naturally lead to completion in 2-3 steps?

---

## Output Requirements

Return a JSON object with EXACTLY this structure:
{{
    "analysis": "Detailed analysis: (1) Why STOP occurred (2) Root cause (3) Which strategy is appropriate (4) How refined task enables 2-3 step completion",
    "strategy": "task_only" or "co_optimize",
    "need_refine": true or false,
    "refined_task": "The optimized task description that enables completion in 2-3 more steps (or empty string if no refinement needed)",
    "step_order": [1, 2, ...],
    "modified_reasonings": {{
        // ONLY include steps whose reasoning needs updates to maintain trajectory coherence after reordering/deletion
        // Key: step number (as string), Value: updated reasoning text
        // Example: {{"3": "Now that we've navigated back, proceed directly to click the search button"}}
    }},
    "completion_estimate": "Specific description of the 1-3 steps needed to complete the refined task from current state"
}}

**Critical Rules**:
1. The refined task MUST be achievable within 2-3 more steps from the current page state.
2. **INTENT over INSTRUCTION**: The refined task must describe the **Goal State** (e.g., "Find the pricing page"), NOT a list of actions (e.g., "Click menu, then click pricing").
3. `step_order` must only contain integers from 1 to {total_steps} (the STOP step should be removed).
4. Each step number can appear at most once in `step_order`.
5. If `strategy` is "task_only", `step_order` must be [{original_order}].
6. For "co_optimize" strategy, ensure modified trajectory is logically coherent.
7. Update `modified_reasonings` for any step whose context changed due to reordering/deletion.

**Focus**: Prioritize enabling the agent to complete the task quickly (2-3 steps) over preserving all trajectory steps. Ensure the task remains a high-level goal, not a low-level script.

RETURN ONLY THE JSON OBJECT WITHOUT ANY COMMENTARY.
\end{promptbox}

\begin{center}
\begin{minipage}{\linewidth}
\captionsetup{hypcap=false}
\captionof{prompt}{Prompt for task-trajectory collaborative refinement.}
\label{prompt:CR}
\end{minipage}
\end{center}

\begin{promptbox}
You are an expert at analyzing and reconstructing web agent trajectories. A trajectory has failed validation and needs to be reconstructed.

## Original High-Level Task
"{high_level_task}"

## Trajectory ({total_steps} steps)
{trajectory_history}

## Validation Issues Detected
{validation_issues}

---

## Your Task

Analyze the trajectory and validation issues, then reconstruct a valid task-trajectory pair by:

1. **Understanding the Context**: What was the agent trying to accomplish? What meaningful progress was made before the issues?

2. **Extracting Valid Steps**: Identify which steps represent meaningful progress and should be kept. You may:
   - **Delete** steps that are redundant, erroneous, or part of a loop
   - **Reorder** steps if needed for logical coherence
   - **Update reasoning** for steps whose context changes due to deletion/reordering
   - You **CANNOT** add new steps - only select from existing ones

3. **Handling Different Issues**:
   - **MISSING_END**: The trajectory lacks a proper NONE (completion) action. If the existing steps accomplish a meaningful sub-goal, rewrite the task to match what was achieved, then indicate we need to append a NONE action.
   - **LOOP_DETECTED**: Remove the redundant loop iterations, keeping only the first occurrence of each unique action in the pattern.
   - **Failed/STOP trajectories**: Extract the meaningful portion of the trajectory that accomplishes a coherent sub-task. Rewrite the task description to match this extracted trajectory.

4. **Ensuring Coherence**: The final trajectory must:
   - Have steps that logically follow each other
   - Match the (possibly rewritten) task description
   - End with a completion state (we'll add NONE if needed)

---

## Output Requirements

Return a JSON object with EXACTLY this structure:
{{
    "analysis": "Detailed analysis of the trajectory issues and your reconstruction strategy",
    "can_reconstruct": true or false,
    "reconstructed_task": "The task description (original or rewritten to match extracted trajectory)",
    "step_order": [list of step numbers to keep, in order, e.g., [1, 2, 4, 5] means delete step 3],
    "modified_reasonings": {{
        // Only include steps whose reasoning needs update
        // Key: original step number (as string), Value: new reasoning
        // Example: {{"4": "After clicking the search button, now viewing results..."}}
    }},
    "needs_none_action": true or false,
    "none_action_value": "Summary/answer for NONE action if needs_none_action is true",
    "reconstruction_summary": "Brief summary of what was changed and why"
}}

**Critical Rules**:
1. `step_order` must only contain integers from 1 to {total_steps}
2. Each step number can appear at most once
3. You can only delete or reorder existing steps, NOT add new ones
4. If `can_reconstruct` is false, explain why in `analysis`
5. The reconstructed trajectory should represent a coherent, completable task
6. If the trajectory is too broken to salvage, set `can_reconstruct` to false

**Example for LOOP_DETECTED**:
If steps 3-4-5-6 form a loop pattern (3-4 repeated), keep only [1, 2, 3, 4] and update reasoning if needed.

**Example for Failed Trajectory**:
If a 10-step trajectory ended in STOP but steps 1-6 accomplished "navigating to product page", rewrite task to "Navigate to the product page" and keep [1, 2, 3, 4, 5, 6].

RETURN ONLY THE JSON OBJECT WITHOUT ANY COMMENTARY.
\end{promptbox}

\begin{center}
\begin{minipage}{\linewidth}
\captionsetup{hypcap=false}
\captionof{prompt}{Prompt for task-trajectory reconstruction in the post-verification.}
\label{prompt:recon}
\end{minipage}
\end{center}

\end{document}